\documentclass{article} 
\usepackage{iclr2027_conference,times}

\usepackage{amsmath,amsfonts,bm}

\def\eqref#1{equation~\ref{#1}}

\def\1{\bm{1}}

\DeclareMathAlphabet{\mathsfit}{\encodingdefault}{\sfdefault}{m}{sl}
\SetMathAlphabet{\mathsfit}{bold}{\encodingdefault}{\sfdefault}{bx}{n}

\usepackage{hyperref}
\usepackage{url}

\usepackage{booktabs}
\usepackage{multirow}
\usepackage{makecell}
\usepackage{graphicx}
\usepackage{wrapfig}
\usepackage{algorithm}
\usepackage{algorithmic}
\usepackage[most]{tcolorbox}
\usepackage[table]{xcolor}
\usepackage{amsmath}
\usepackage{amssymb}
\usepackage{amsthm}

\providecommand{\se}[1]{\ensuremath{_{\pm#1}}}
\newcommand{\Ddev}{\mathcal{D}_{\mathrm{dev}}}

\newtcolorbox{promptbox}{
  colback=gray!5,
  colframe=gray!60,
  boxrule=0.5pt,
  arc=2pt,
  left=6pt,
  right=6pt,
  top=6pt,
  bottom=6pt,
  fontupper=\ttfamily\small,
  breakable
}

\title{
Divide, Diagnose, and Compose: Optimizing Agentic Pipelines and Harnesses
}
\author{
\parbox[t]{\textwidth}{\centering 
Divij Handa\textsuperscript{1} \thanks{Work done during internship at Adobe} \quad
Christopher Tensmeyer\textsuperscript{2} \quad
Jihyung Kil\textsuperscript{2} \quad
Vlad I. Morariu\textsuperscript{2}}
\\
\parbox[t]{\textwidth}{\centering \textsuperscript{1} Arizona State University \quad
\textsuperscript{2} Adobe}
\\
\parbox[t]{\textwidth}{\centering \texttt{dhanda@asu.edu, \{tensmeye,jkil,morariu\}@adobe.com}}
}

\iclrfinalcopy

\begin{document}
\maketitle

\fancyhead{}
\renewcommand{\headrulewidth}{0pt}

\begin{abstract}

Building agentic pipelines that achieve high task performance while keeping cost and latency low requires jointly optimizing heterogeneous and coupled design choices, including prompts, topology, model selection, and tool policies. Manually searching this combinatorial design space is expensive and time-consuming, while existing automated optimizers typically operate within a predefined subset of this space, modify the pipeline monolithically, or provide little structured evidence for pipeline changes. To mitigate these challenges, we introduce the Decomposed Credit-Assignment Framework (\textsc{DeCAF}), which adapts the design space to the given benchmark and pipeline, optimizes each dimension separately, and generates reasoning traces for every decision. Given a task specification, benchmark examples, and a seed pipeline, \textsc{DeCAF} identifies relevant optimization dimensions and generates diverse experimental hypotheses for each. Independent coding-agent sessions then optimize each dimension under a restricted edit scope, jointly evaluating performance, inference cost, and latency while recording all hypotheses, edits, and outcomes of each experiment. A final synthesis session combines compatible findings across all dimensions and uses a component-level trajectory evaluator, which derives role-specific rubrics from each agent's instructions, to localize underperforming components and guide cross-dimension revisions. Across a diverse suite of agentic benchmarks, \textsc{DeCAF} improves average task performance by 2.58 percentage points, while reducing inference cost by 4.30\% and end-to-end latency by 42.45\% compared to the best overall baseline.

\end{abstract}

\section{Introduction}

\begin{wrapfigure}{r}{0.51\textwidth}
    \centering
    \vspace{-30pt}
    \includegraphics[width=\linewidth]{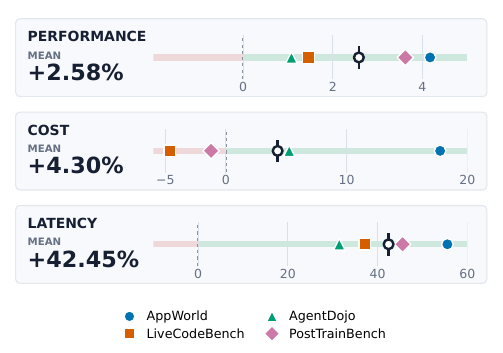}
    \vspace{-20pt}
    \caption{Improvements of \textsc{DeCAF} over Meta-Harness (best baseline) across all benchmarks.}
    \label{fig:teaser-res}
\end{wrapfigure}

\begin{figure*}[ht]
    \centering
    \includegraphics[width=0.98\linewidth]{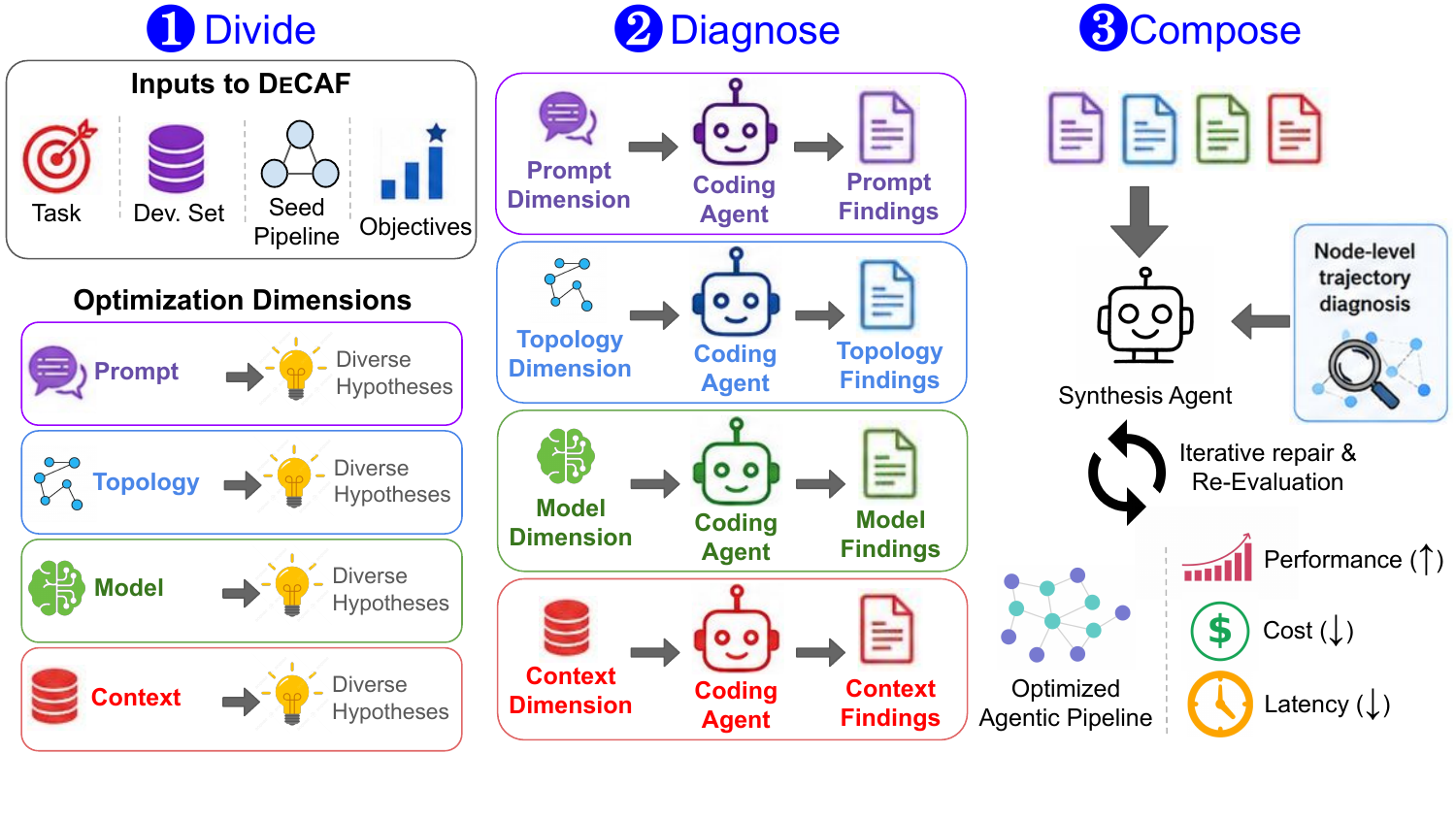}
    \vspace{-20pt}
    \caption{
        Given task details, development examples, seed agentic pipeline, and objectives, \textsc{DeCAF} first identifies task-relevant optimization dimensions and generates diverse experimental hypotheses for each (\textit{Divide}). Independent coding-agent sessions then optimize copies of the seed pipeline under dimension-specific edit constraints, producing Pareto candidates, execution trajectories, and structured findings (\textit{Diagnose}). Finally, a synthesis optimizer combines compatible dimension-level findings and performs targeted cross-dimension revisions to produce a unified pipeline on the performance–cost–latency frontier with the help of a component or node-level diagnosis tool (\textit{Compose}).
    }
    \label{fig:teaser-overview}
\end{figure*}

Large language models (LLMs) are increasingly deployed not as isolated models but as components of larger \textit{agentic pipelines} and \textit{harnesses} that combine multiple model calls with tools, memory, retrieval, verification, and control flow. The behavior of such pipelines depends on substantially more than just the choice of a base model. Developers must decide what instructions each component receives, how components exchange information, which models serve which roles, how much inference-time computation each call receives, what context is retained or discarded, and when intermediate outputs should be verified or revised. These design choices are coupled: a cheaper model may demand stronger scaffolding, and an added verifier increases both cost and latency. Therefore, manually tuning an agentic pipeline for competing objectives such as performance, cost, and latency is expensive and requires substantial task-specific expertise \citep{zhou2025multi, hua2026agentopt}. As a result, a growing body of work shows that autonomously optimizing all prompts, demonstrations, workflow structure, model selection, context, or agent roles can each produce meaningful gains in multi-LLM agentic pipelines \citep{wu2026optimas, saeidi2026fama, zhang2025aflow, agrawal2025gepa, opsahl2024optimizing}.

These techniques, however, exhibit one or more of the following limitations. First, several optimizers are constrained by design to a narrow set of dimensions. For example, GEPA \citep{agrawal2025gepa} optimizes just the prompts of all agents, while AFlow \citep{zhang2025aflow} works by optimizing the topology of the pipeline. Optimizing multiple dimensions jointly has been shown to outperform optimizing just a single dimension \citep{zhou2025multi}. Second, LLMs tend to produce solutions that lack diversity \citep{handa2026guidedsampling, brown2024large}, a weakness that is amplified when they act as optimizers and must propose genuinely distinct candidate edits to a pipeline \citep{lee2026meta}. Third, several optimizers proceed by trial and error without producing a reasoning trace, which makes the resulting changes difficult to understand, audit, or build upon \citep{zhang2025multi}. Finally, recent work argues that measuring each agent's individual contribution provides a more informative reward than a single final binary outcome score \citep{wang2026maspo}. Such fine-grained credit reveals which component is responsible for a failed or successful trajectory. To address these challenges, we introduce the Decomposed Credit-Assignment Framework (\textsc{DeCAF}) for optimizing multi-agent pipelines and harnesses. Figure~\ref{fig:teaser-overview} illustrates its overall workflow.

\textsc{DeCAF} begins by reading the task description, a few development examples, the seed agentic pipeline to be optimized, and the objectives (or metrics) it has to achieve. We treat each pipeline as a computational graph $G=(V,E)$, where each node $v \in V$ is an executable component of the pipeline, such as an LLM agent initialized with a particular role. Conditioned on the benchmark and seed pipeline information, \textsc{DeCAF} starts by proposing the dimensions along which the pipeline can be optimized. Then for each dimension it iteratively generates diverse experiments \citep{handa2026guidedsampling} to improve exploration of ideas to test. Each dimension's experiments are executed by an independent coding-agent session that is restricted to edits within that dimension's scope. The findings of all dimension-specific sessions are recorded and passed to a final stage, in which a single coding agent optimizes the whole pipeline using the accumulated evidence. This synthesis agent is additionally given a \texttt{node\_eval} tool, which reads trajectories generated by the current pipeline and produces a detailed score for each component, localizing the agents most in need of revision.

\textsc{DeCAF}’s main contribution is the way it organizes open-ended search for optimization. Rather than restricting optimization to a predefined design axis or tasking a single optimizer to modify the full pipeline monolithically, \textsc{DeCAF} (i) adaptively decomposes the search space into task and pipeline-specific dimensions that are explored in independent coding-agent sessions, (ii) forces exploration within each dimension by generating diverse experiment hypothesis, and (iii) recomposes all dimension-level findings in a final synthesis stage using component-level trajectory feedback. Together, these mechanisms make pipeline optimization more structured, traceable, and amenable to localized diagnosis. Across held-out evaluations on a diverse suite of agentic benchmarks, \textsc{DeCAF} improves the performance-cost-latency frontier over baselines. Compared to the strongest overall baseline, Meta-Harness \citep{lee2026meta}, it improves task performance by an average of 2.58pp, while reducing per-task inference cost by 4.30\% and end-to-end latency by 42.45\% (Figure~\ref{fig:teaser-res}).

\section{Related Work}

\paragraph{Automated Optimization of Agentic Pipelines and Harnesses.}
A large body of work automates parts of the design space. OPRO \citep{yang2024large} frames instruction search as optimization where the model is shown its previously proposed instructions alongside their measured scores and asked to write one that will score higher. MIPRO \citep{opsahl2024optimizing} jointly searches over the instructions and few-shot demonstrations of every module in a multi-stage program. TextGrad \citep{yuksekgonul2024textgrad} builds an analogy to backpropagation: an LLM writes a natural-language critique of each intermediate output with respect to the final objective, and the critique propagates backward to revise the prompt that produced it. GEPA \citep{agrawal2025gepa} reflects in natural language on full execution traces to propose targeted edits and retains any candidate that is best on at least one training instance. MASPO \citep{wang2026maspo} brings prompt optimization to multi-agent systems, scoring each agent's prompt not only by the quality of its own output but also by its effect on the agents downstream. ADAS \citep{hu2025automated} tasks a meta-agent with writing complete agent designs as executable programs, evaluating each and conditioning on the archive of past designs to invent the next. AFlow \citep{zhang2025aflow} likewise represents workflows as code, but explores modifications with Monte-Carlo tree search, letting execution feedback decide which branches deserve expansion. MASS \citep{zhou2025multi} demonstrates that prompts and topology are best optimized together, alternating between per-agent prompt tuning, topology search over a pruned space of influential configurations, and a final global prompt pass. MaAS \citep{zhang2025multi} replaces the single best architecture with an ``agentic supernet'', a learned distribution over architectures from which a query-conditioned pipeline is sampled, so easy queries receive cheap systems and hard ones receive more compute. Meta-Harness \citep{lee2026meta} broadens the optimization space by asking a single coding agent to optimize the entire pipeline or harness. \textsc{DeCAF} on the other hand splits and optimizes the entire space into independent dimensions and then synthesizes the findings.

\textbf{Failure Localization and Credit Assignment.}\quad
An end-to-end task reward indicates whether a complete pipeline succeeded, but often provides little information about where a multi-component execution failed. \citet{zhang2025agent} formalize automated failure attribution in LLM-based multi-agent systems as identifying both the agent and the decisive step associated with a failure. Recent work has introduced several forms of credit assignment for training and optimizing agents. Agent Lightning \citep{luo2025agent} decouples agent execution from reinforcement-learning infrastructure and includes a credit-assignment module that transforms arbitrary agent trajectories into training transitions. \citet{li2026unifying} decompose credit along temporal and structural axes, identifying weak interaction rounds and roles before refining role prompts and aggregation protocols through verbalized block-coordinate descent. MASPO \citep{wang2026maspo} evaluates multi-agent prompts using local validity, lookahead potential for successor agents, and global alignment with the final system outcome. In our work, the \texttt{node\_eval} tool used in the final stage of \textsc{DeCAF} generates a fine-grained score for each agent in the pipeline. Because this evaluation is based primarily on observed trajectories, \textsc{DeCAF} treats them as evidence for localizing \textit{likely} underperforming components rather than as exact causal estimates of each node's contribution.

\textbf{Coding Agents for Open-Ended System Optimization.}\quad
Software-engineering agents can implement nontrivial changes to real repositories \citep{yang2024swe, zhang2025buildbench}, and self-improving systems such as the Darwin G\"odel Machine \citep{zhang2026darwin} use coding agents to rewrite their own agentic code. Meta-Harness \citep{lee2026meta} similarly formulates harness engineering as an outer-loop code-search problem. Its coding-agent proposer can inspect the harness source code, scores, and execution traces of prior candidates through a filesystem, and can then modify and evaluate the complete harness. Similarly, \textsc{DeCAF} uses coding agents as optimization workers, but differs in how their search is organized. A hypothesis generator is used before the coding agent to boost exploration and a node-level diagnostic tool is used for better credit assignment.

\section{DeCAF}
\subsection{Problem Formulation}
\label{sec:decaf-formulation}

Let $P_0=(G,\Theta)$ denote a seed agentic pipeline, where $G=(V,E)$ is its computational graph and $\Theta$ contains configurable choices such as prompts, model assignments, inference parameters, context policies, tools, and control flow. Executing a pipeline $P$ on input $x$ produces an output $\hat{y}$ and a trajectory $\tau$ containing intermediate node inputs, outputs, tool calls, and system state. For a development set $\mathcal{D}_{\mathrm{dev}}=\{x_i\}_{i=1}^{N}$, we associate each pipeline with the objective vector
{\small
\begin{equation}
    \mathbf{J}(P)
    =
    \frac{1}{N}
    \sum_{i=1}^{N}
    \Bigl(
        q(P,x_i),\,
        -c(P,x_i),\,
        -\ell(P,x_i)
    \Bigr),
    \label{eq:objective-vector}
\end{equation}
}
where $q$, $c$, and $\ell$ measure task performance, inference cost, and end-to-end latency, respectively. A candidate dominates another if it achieves no lower task performance, no greater cost or latency, and strictly improves at least one objective. While objectives can be folded into a single scalar by weighted combination, e.g., $0.8\times q(P,x_i) - 0.1\times c(P,x_i) - 0.1\times\ell(P,x_i)$, this scalarization commits to weights before the attainable trade-offs are known, yields one operating point per run, and has well-known failure modes (e.g., non-convex regions of the frontier are unreachable by any linear weighting) \citep{deb2002fast}. Therefore, rather than collapsing these objectives into a fixed weighted scalar, \textsc{DeCAF} maintains non-dominated candidates, preserving operating points with different deployment trade-offs. The held-out test set is not exposed during any part of \textsc{DeCAF}.

\subsection{Divide: Task-Adaptive Dimensions and Experiments}
\label{sec:decaf-divide}

\textsc{DeCAF} first inspects the task description, a small subset of development examples, the implementation of the seed pipeline, and the objectives as defined in \textsection\ref{sec:decaf-formulation}. Conditioned on few-shot decomposition examples, a dimension planner proposes task-relevant optimization dimensions $D_1,\ldots,D_K$. Each dimension is represented as $D_k=(\mathcal{M}_k,\mathcal{C}_k,\rho_k)$, where $\mathcal{M}_k$ specifies the editable pipeline components, $\mathcal{C}_k$ specifies scope and validity constraints, and $\rho_k$ records why the dimension is relevant to the task. Candidate dimensions include prompt optimization, workflow topology, model and inference configuration, context management, tool policy, memory, and agent skills. The proposed dimensions are benchmark-and-pipeline-dependent rather than fixed across all tasks. For each dimension $D_k$, an experiment generator proposes a set of hypotheses $\mathcal{H}_k=\{h_{k,1},\ldots,h_{k,n_k}\}$. Each hypothesis describes a distinct configuration to investigate rather than a complete implementation. For example, a topology dimension may consider adding a verifier, introducing a planner--executor structure, or adding a targeted repair loop; a model-configuration dimension may consider role-specific model assignments or different reasoning effort.

To maximize exploration of each dimension, each hypothesis is generated iteratively. At each iteration, the generator is conditioned on previously proposed hypotheses and instructed to produce a distinct experimental direction. Inspired by GuidedSampling \citep{handa2026guidedsampling}, this iterative procedure reduces evaluation budget spent on superficial variations of the same idea. Each hypothesis records its motivation, affected components, and expected effects on performance, cost, and latency.

\subsection{Diagnose: Constrained Dimension-Level Optimization}
\label{sec:decaf-diagnose}

For every dimension $D_k$, \textsc{DeCAF} creates a copy of the seed pipeline and launches a fresh coding-agent session for optimization. The session receives the task specification, development data, dimension definition, and hypotheses $\mathcal{H}_k$. It has a primary editable scope $\mathcal{M}_k$ and is instructed to focus its search on dimension $D_k$. However, supporting changes are allowed. For example, when modifying topology, new prompts need to be written for the agents. The `topology' still remains the primary optimization dimension, while `prompt' becomes the secondary or supportive dimension. The coding agent iteratively selects or adapts a hypothesis, implements the corresponding changes, and evaluates the candidate. Each dimension maintains a non-dominated archive
\begin{equation}
    \mathcal{A}_k
    \leftarrow
    \operatorname{ND}
    \left(
        \mathcal{A}_k \cup \{P_{k,j}\}
    \right),
    \label{eq:pareto-update}
\end{equation}
where $\operatorname{ND}$ removes candidates dominated by another member of the archive. A candidate need not strictly dominate the seed to be retained. For example, a performance gain at a modest cost increase may represent a useful operating point on the observed frontier. Every attempted experiment is also recorded in a \texttt{FINDINGS.md} artifact. The record includes the motivating hypothesis, affected components, implemented change, measured objectives, whether the candidate was retained, observed regressions, and conclusions for later optimization. Failed experiments are preserved so that the final optimizer can avoid repeating ineffective changes and can reason about benchmark-specific trade-offs. Because all dimension-level sessions begin from the same seed and vary only their assigned design dimension, their outcomes provide controlled, conditional evidence about which classes of interventions are effective around $P_0$. To reduce the wall-clock optimization time, the sessions run in parallel. Each dimension returns its archive and findings, $(\mathcal{A}_k,\mathcal{F}_k)$, to the final stage.

\subsection{Compose: Node-Guided Evidence Synthesis}

Independently successful changes need not remain successful when combined. For example, adding a verifier may require corresponding prompt changes, while context compression may interact with a new model assignment. \textsc{DeCAF} therefore launches a final coding-agent session that receives the seed pipeline, all dimension-level archives and findings, and the task objectives. Unlike the dimension-level optimizers, this final synthesis agent may modify the complete pipeline. To help the synthesis agent understand why a pipeline succeeds or fails, we provide a diagnostic tool called \texttt{node\_eval}. Instead of looking only at the final task score, \texttt{node\_eval} examines each node across multiple development trajectories. It evaluates three main aspects of a node: \emph{correctness}, whether the node makes a sound contribution toward the user's goal; \emph{compliance}, whether it follows the instructions in its own system prompt; and \emph{downstream utilization}, whether its output is passed to and used by later components. Evaluating each node across multiple development trajectories reduces sensitivity to isolated failures and helps reveal recurring behavior \citep{li2026unifying}.

For node $v$ and evaluation criterion $r$, let $s_{v,r}^{(\tau)}$ denote the score assigned on development trajectory $\tau$. Some criteria may not apply to every trajectory, or the available trace may not contain enough evidence to score them. We therefore define $\mathcal{T}_{v,r}^{\mathrm{scored}}$ as the set of trajectories for which criterion $r$ receives a numeric score, and summarize the node's behavior as
\begin{equation}
    \bar{s}_{v,r}
    =
    \frac{1}{|\mathcal{T}_{v,r}^{\mathrm{scored}}|}
    \sum_{\tau \in \mathcal{T}_{v,r}^{\mathrm{scored}}}
    s_{v,r}^{(\tau)}.
    \label{eq:node_eval}
\end{equation}
In addition to these scores, \texttt{node\_eval} returns evidence and natural-language descriptions of recurring problems. By examining the node output together with the information delivered to downstream components and the final pipeline result, it can distinguish likely problems in the source node, its consumer, or the handoff between them. These diagnoses are based on observed trajectories, so we use them to localize likely failures rather than to claim exact causal responsibility.

The synthesis agent uses these diagnostics together with the successful and unsuccessful findings from the dimension-level searches. It tests combinations of promising changes and re-evaluates every composed pipeline end-to-end on $D_{\mathrm{dev}}$. When a combination regresses, \texttt{node\_eval} can help identify where the interaction failed and guide a targeted repair, such as changing a prompt or adjusting the information passed between stages. Importantly, \texttt{node\_eval} is only a diagnostic tool: pipeline changes are still accepted or rejected using the task-performance, cost, and latency objectives. The synthesis stage maintains a final non-dominated archive using the same update rule as Eq.~\ref{eq:pareto-update}. A final pipeline and harness is selected via the optimizer LLM from this archive using development-set evidence and is then evaluated on the held-out test set.

\section{Experiment Setup}
\paragraph{Benchmarks and Data Splits.}
We evaluate \textsc{DeCAF} on four benchmarks: AppWorld \citep{trivedi2024appworld}, LiveCodeBench \citep{jain2025livecodebench}, AgentDojo \citep{debenedetti2024agentdojo}, and PostTrainBench v1.1 \citep{rank2026posttrainbench}. For AppWorld, we optimize on the official \texttt{dev} split and evaluate on \texttt{test\_normal}. For LiveCodeBench, we use the code-generation setting and create a split from release\_v6, dropping problems before 2024-10-01. We then use 20\% (51 instances) of the data as development set, while 80\% (206 instances) is used as the held-out test set. We also remove the ``Easy'' split, since it was saturated. For AgentDojo, we create development and test splits at the user-task level. We use 20 user tasks (60 episodes) for development and 77 user tasks (231 episodes) for testing, with each task evaluated once benignly and under two attacks. For PostTrainBench, we construct development and test settings across both target base models and benchmark.
{\small
\[
    \mathcal{M}_{\mathrm{dev}}
    =
    \{\text{Qwen3-1.7B},\text{SmolLM3-3B}\},
    \quad
    \mathcal{B}_{\mathrm{dev}}
    =
    \{\text{GSM8K},\text{GPQA Main},\text{AIME 2025}\},
\]
\[
    \mathcal{M}_{\mathrm{test}}
    =
    \{\text{Qwen3-4B},\text{Gemma-3-4B}\},
    \quad
        \mathcal{B}_{\mathrm{test}}
    =
    \{\text{HumanEval},\text{AIME 2026},\text{BFCL}\}
\]
}
Thus, the final evaluation tests whether an optimized post-training pipeline transfers to both unseen base models and unseen task domains. For BFCL, we use the full BFCL-V3 evaluation suite \citep{patil2025berkeley} rather than the smaller subset distributed with PostTrainBench v1.1.

\textbf{Seed pipelines and Models.}\quad
Each optimization begins from a fixed seed pipeline. For AppWorld, we use a single ReAct coding harness, backed by Claude-Sonnet-4.6 at \texttt{medium} effort, with other Claude models as choices for optimization. For LiveCodeBench, we use a sequential generator--critic--refiner pipeline in which all three nodes are initialized with GPT-5.4-nano, and GPT-5.4 family of models given for optimization. For AgentDojo, we use the seed harness with a single-stage ReAct agent using Qwen-3.5-397b-a17b-FP8 model, and Gemma \citep{team2026gemma} and Qwen \citep{team2026qwen3} family of models are available as optimization choices. For PostTrainBench, a single Claude Code post-training agent backed by Claude-Sonnet-5 at \texttt{high} effort is used as the seed. Opus and Haiku variants are given as choices for optimization. \textsc{DeCAF} uses Claude Code as the coding agent and uses \texttt{/goal} mechanism\footnote{\url{https://code.claude.com/docs/en/goal}} to maintain the optimization objective. Every optimizer is run twice and the final table reports the average and standard error across both runs.

\textbf{Baselines.}\quad
We compare \textsc{DeCAF} against the unoptimized seed pipeline and three automated optimizers. We additionally evaluate the unoptimized seed pipeline using backbone models from different capability tiers. These baselines measure whether the gains from optimization can be reproduced by simply replacing the underlying model with a more capable and expensive alternative. GEPA \citep{agrawal2025gepa} performs reflective evolutionary optimization over the prompts of all LLM nodes. We initialize every GEPA optimization with the \texttt{heavy} budget \footnote{\url{https://dspy.ai/api/optimizers/GEPA/overview/}} for a thorough optimization. MASPO \citep{wang2026maspo} jointly optimizes node prompts using local validity, downstream usefulness, and global task feedback. Meta-Harness \citep{lee2026meta} gives a coding-agent proposer access to the complete history of previous harnesses, their scores, source code, and raw execution traces through a filesystem, and lets the agent decide what to inspect and what to modify. We initialize Meta-Harness with $N=\# Dimensions$ and $k=\# Experiments/Dimension$ to keep the experiment budget similar to \textsc{DeCAF}.
All optimizers use Claude-Sonnet-5 at \texttt{high} effort. For methods requiring a scalar candidate score, we normalize all objectives relative to the seed pipeline:
{\small
\begin{equation}
\begin{split}
    U(P)
    ={}&
    w_Q \cdot Q(P)
    +
    w_C
    \frac{C(P_0)-C(P)}{C(P_0)}
    +
    w_L
    \frac{L(P_0)-L(P)}{L(P_0)},
    \label{eq:baseline-scalarization}
\end{split}
\end{equation}
}
where $(w_Q,w_C,w_L)= (0.8,0.1,0.1)$ is fixed before test evaluation and shared across benchmarks. \textsc{DeCAF} and Meta-Harness receive the three objectives separately and maintain non-dominated candidates rather than optimizing this scalarization. The optimizing LLM is asked to return a single pipeline that they believe is best given the three criteria.

\section{Results}

\begin{table}[t]
\caption{
Task scores (\% $\pm$ standard-error), cost-per-instance, latency-per-instance, one-time cost of optimization, and wall-clock time (hh:mm) of all optimizers. Each row is an average of two runs.
}
\label{tab:main_results}
\centering
\footnotesize
\setlength{\tabcolsep}{3pt}
\begin{tabular}{@{}lcccrr@{}}
\toprule
& & \multicolumn{2}{c}{\textbf{Deployment / instance}}
  & \multicolumn{2}{c@{}}{\textbf{Optimization}} \\
\cmidrule(lr){3-4} \cmidrule(l){5-6}
\textbf{Method}
    & \textbf{Task score} $\uparrow$
    & \textbf{Cost (\$)} $\downarrow$
    & \textbf{Latency (s)} $\downarrow$
    & \textbf{Cost (\$)} $\downarrow$
    & \textbf{Time} $\downarrow$ \\

\midrule

\multicolumn{6}{@{}l}{\emph{AppWorld (TGC)}} \\
\quad Seed pipeline (Sonnet-4.6-Medium)  & 78.17\se{1.21} & 0.2242 & 55.1  & --     & --    \\
\quad Stronger backbone (Opus-4.8-High)  & 87.20\se{0.89} & 0.4273 & 114.3 & --     & --    \\
\quad GEPA         & 74.40\se{1.19} & 0.2901 & 58.8 & 279.04 & 14:57 \\
\quad MASPO        & 78.17\se{1.21} & 0.2242 & 55.1 & 103.57 & 02:06 \\
\quad Meta-Harness & 80.06\se{0.89} & 0.2237 & 73.8 & 166.91 & 07:58 \\
\rowcolor{gray!10}%
\quad \textbf{\textsc{DeCAF} (Ours)} & 84.23\se{0.89} & 0.1840 & 32.8 & 191.86 & 04:28 \\

\midrule

\multicolumn{6}{@{}l}{\emph{LiveCodeBench (pass@1)}} \\
\quad Seed pipeline (GPT-5.4-nano) & 47.33\se{1.13} & 0.0051 & 29.3 & --    & --    \\
\quad Stronger backbone (GPT-5.4)  & 55.34\se{0.56} & 0.0236 & 12.5 & --    & --    \\
\quad GEPA         & 48.30\se{1.70} & 0.0055 & 37.2 & 24.60 & 09:31 \\
\quad MASPO        & 48.30\se{0.24} & 0.0052 & 26.4 & 74.21 & 02:48 \\
\quad Meta-Harness & 51.70\se{1.21} & 0.0065 & 44.4 & 57.22 & 07:04 \\
\rowcolor{gray!10}%
\quad \textbf{\textsc{DeCAF} (Ours)} & 53.16\se{0.73} & 0.0068 & 27.9 & 81.88 & 04:37 \\

\midrule

\multicolumn{6}{@{}l}{\emph{AgentDojo}} \\
\quad Seed pipeline (Qwen-3.5-397b-a17b-FP8) & 66.71\se{0.70} & 0.0163 & 16.45 & --    & --    \\
\quad Stronger backbone (Gemma-4-31B-it)     & 80.93\se{0.35} & 0.0060 & 49.33 & --    & --    \\
\quad GEPA         & 83.76\se{0.51} & 0.0137 & 24.23 & 32.92 & 04:17 \\
\quad MASPO        & 87.76\se{0.54} & 0.0147 & 36.66 & 21.05 & 01:14 \\
\quad Meta-Harness & 86.69\se{0.24} & 0.0133 & 37.40 & 28.74 & 07:11 \\
\rowcolor{gray!10}%
\quad \textbf{\textsc{DeCAF} (Ours)} & 87.77\se{0.45} & 0.0126 & 25.63 & 26.48 & 04:43 \\

\midrule

\emph{PostTrainBench} & & & \emph{(hh:mm)} & \\
\quad Seed pipeline (Sonnet-5-High) & 33.56\se{1.55} & 10.07 & 08:01 & -- & --    \\
\quad Stronger backbone (Opus-4.8-Max)  & 40.54\se{0.93} & 20.20 & 08:17 & -- & --    \\
\quad GEPA         & 33.56\se{1.55} & 10.07 & 08:01 & 154.26 & 20:41 \\
\quad MASPO        & 33.56\se{1.55} & 10.07 & 08:01 & 86.10 & 06:21 \\
\quad Meta-Harness & 31.97\se{2.77} & 9.90 & 07:04 & 145.63 & 42:46 \\
\rowcolor{gray!10}%
\quad \textbf{\textsc{DeCAF} (Ours)} & 35.59\se{1.00} & 10.02 & 03:51 & 177.25 & 43:52 \\

\bottomrule
\end{tabular}
\end{table}

\subsection{End-to-End Performance--Cost--Latency Trade-offs}
\label{sec:results-main}

Table~\ref{tab:main_results} compares the final pipelines produced by each optimizer. Across the four benchmarks, \textsc{DeCAF} improves task performance over Meta-Harness by an average of $2.58$ percentage points while reducing inference cost by $4.30\%$ and latency by $42.45\%$. The largest gains occur on AppWorld, where \textsc{DeCAF} improves TGC by $4.17pp$ while reducing cost by $17.8\%$ and latency by $55.6\%$. On LiveCodeBench and AgentDojo, it improves performance by $1.46pp$ and $1.08pp$ respectively, with substantial latency reductions. LiveCodeBench incurs a small increase in inference cost (\$0.0003, $4.6\%$). On PostTrainBench, \textsc{DeCAF} improves the task score over Meta-Harness while roughly halving deployment latency, although optimization is substantially more expensive because GPU constraints prevent parallel dimension-level search. Stronger unoptimized backbones can exceed \textsc{DeCAF} in raw task performance on some benchmarks, but generally require substantially greater deployment cost. MASPO on AppWorld and GEPA and MASPO on PostTrainBench regress during their optimization runs and hence return the seed pipeline as is. Detailed benchmark-specific optimizations are provided in Appendix~\ref{app:full-results}.

\begin{figure*}[ht]
    \centering
    \includegraphics[width=0.9\linewidth]{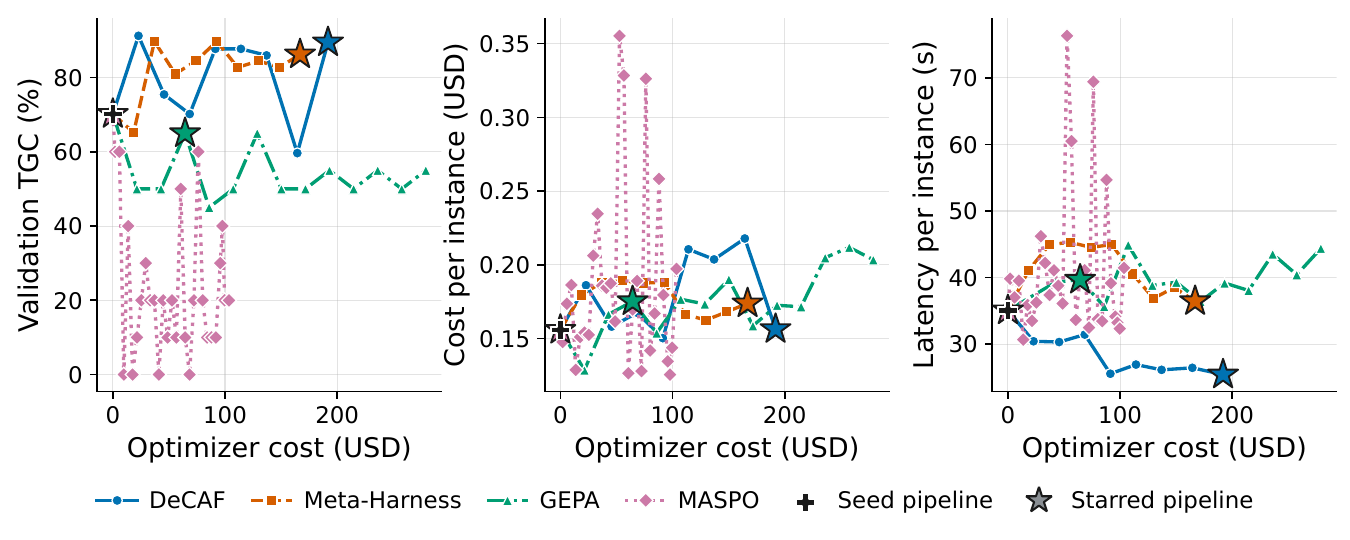}
    \vspace{-15pt}
    \caption{
        AppWorld optimization trajectories as a function of cumulative optimizer cost. Each point corresponds to an evaluated validation candidate; the panels report TGC, per-instance inference cost, and latency, respectively, while stars denote the pipeline ultimately selected by each optimizer. Trajectories need not be monotonic as they show individual evaluated candidates not the best so far.
    }
    \label{fig:appworld-progress}
    \vspace{-10pt}
\end{figure*}

\textbf{Optimization dynamics.}\quad
Figure~\ref{fig:appworld-progress} plots each optimizer against cumulative expenditure. \textsc{DeCAF} reaches high validation TGC early in the search while exploring substantially lower-latency configurations. Its trajectory is non-monotonic because independent dimension-level searches explore different regions of the design space. The final synthesis evaluations combine these findings and produce the selected pipeline, which slightly exceeds Meta-Harness in validation TGC while substantially reducing deployment cost and latency. Even though both Meta-Harness and \textsc{DeCAF} run 9 experiments, \textsc{DeCAF} is able to exploit the solution space more effectively. Meta-Harness TGC and cost start saturating at just 22.2\% of its budget while the latency drops and saturates at 77.8\%.

\subsection{Optimization Overhead of DeCAF}
\label{sec:results-overhead}

Table~\ref{tab:main_results} reports the one-time cost and wall-clock time required to run each optimizer. Since \textsc{DeCAF} decomposes the search space into independent dimensions, coding agents for each dimension can run in parallel. This reduces the wall-clock time of \textsc{DeCAF} significantly compared to Meta-Harness, where experiments run sequentially. The only exception being PostTrainBench, where we ran each dimension sequentially due to the bottleneck on the GPU resources. This makes \textsc{DeCAF} $27.6\%$ faster on average ($37.65\%$ faster if we remove PostTrainBench). However, this decomposition and better exploration comes at a cost, it increases the total cost of optimization. On average, \textsc{DeCAF} is $17.98\%$ more expensive than Meta-Harness. There are two major contributions of this increased cost. First, the optimizer is forced to explore performance using the stronger (and generally more expensive) models. This increases the cost of evaluation of the pipeline on the validation set. Second, the \texttt{node\_eval} tool makes up roughly $6\%$ of the optimization cost of \textsc{DeCAF}. This tool takes in multiple trajectories of the agent which increases the input token cost. Due to the heavy cost of this tool, it is only given to the final synthesizer coding agent, not the dimension-level coding agents.

\subsection{Ablation Study}
\label{sec:results-ablation}

To identify the contribution of each part of \textsc{DeCAF}, we perform ablations on AppWorld and LiveCodeBench benchmarks. We consider the following variants: \textbf{No Dimensions} removes task-adaptive decomposition while preserving the number of coding-agent optimizers. A global experiment generator proposes the same total number of hypotheses as full \textsc{DeCAF} without conditioning on a particular dimension. The hypotheses are divided evenly among four independent coding-agent sessions, each tasked with optimizing the seed pipeline and is given unrestricted edit access. Their resulting pipelines and findings are then provided to the synthesis agent. \textbf{Single Shared Session} retains the dimensions and experiment proposals produced by full \textsc{DeCAF}, but replaces the independent dimension-specific coding-agent sessions with a single coding-agent session. The single agent optimizes the seed-pipeline for all dimensions. This tests whether independent optimization contexts reduce interference when searching heterogeneous parts of the design space. \textbf{No Experiment Proposals} retains the task-adaptive dimensions and independent coding-agent sessions but removes the diverse experiment hypothesis. Each coding agent instead is tasked with generating its own experiments during optimizing its assigned dimension. \textbf{No \texttt{node\_eval}} uses the same final synthesis coding agent, but removes access to the component-level trajectory diagnostics. The synthesis agent can still inspect ordinary execution traces and evaluation results by itself. \textbf{No Synthesis} retains the complete Divide and Diagnose stages of \textsc{DeCAF} but performs no cross-dimensional optimization. The final pipeline is constructed simply as the combination of the dimension-specific optimized pipeline. This measures how much improvement arises from combining and repairing dimension-level findings. \textbf{Dimension Best} evaluates the pipeline by all dimensions on the validation set and returns the best one skipping the entire synthesis stage. It differs from No Synthesis, as it doesn't combine the findings across dimensions.

Table~\ref{tab:ablation} reports the results of all ablations. Across both benchmarks, a big drop is observed on the \textit{No Dimensions} variant, where we remove the dimension proposer, and directly ask the parallel coding agents to optimize the pipeline. Upon inspecting the trajectories of the optimizers we found that roughly 41.67\% of the budget was spent on context engineering, followed by spending 16.67\% of the budget in Harness Optimization. This left very little budget to effectively explore other dimensions like Prompt Optimization, Topology Variations, and Different Agent Configurations. Similarly in the \textit{No Experiment Proposal} variant, both benchmarks regressed significantly. Upon further inspection, we found that the optimizer either stops after finding a single solution better than the seed pipeline, or it tries to work out a single hypothesis without exploring alternatives. Both these variants highlight the limitations of existing models to effectively explore.

The rest of the variations help answer the question: ``Are the dimensions truly independent of each other and whether they interact positively or not?''. Both the \textit{No \texttt{node\_eval}} and \textit{No Synthesis} variations show little degradation in AppWorld but degrades by up to 8.35pp on LiveCodeBench. This highlights that optimizations made by individual dimensions within AppWorld can simply be added up without finetuning the pipeline, whereas in LiveCodeBench these optimizations are destructive to each other. This highlights why the final synthesis stage of \textsc{DeCAF} is really important, especially the \texttt{node\_eval} tool. The \textit{Single Shared Session} variation further highlights this distinction. On AppWorld, where dimensions interact positively with each other, having a single coding agent degrades the performance. On LiveCodeBench however, the performance drop is insignificant since all the dimension optimizations need to be tested together. Finally, \textit{Dimension Best} shows why different dimensions are important. On AppWorld, a simple model scaling gives a better result (+2.97pp) but with a much higher cost (+132.23\%) and latency (+248.49\%). On LiveCodeBench, context engineered pipeline highlighting potential errors was selected. Since the dimension was optimized independently there weren't any conflicts from other dimensions leading only to slight decrease in performance, while cost remained the same, and latency increased slightly. Since it's impossible to know beforehand whether the dimensions will interact positively or negatively, we believe the three stages of \textsc{DeCAF} (Divide, Diagnose, and Compose) helps recover performance in both scenarios.

\begin{table*}[t]
\centering
\caption{
Ablation study on the AppWorld and LiveCodeBench benchmarks. $\Delta$ denotes the absolute change in the benchmark-specific task performance metric relative to the full \textsc{DeCAF} method. Per-instance columns report test-time deployment metrics, while optimizer columns report one-time optimization overhead.
}
\label{tab:ablation}
\footnotesize
\setlength{\tabcolsep}{6pt}
\renewcommand{\arraystretch}{1.08}

\begin{tabular}{@{}l
                cc
                cc
                cc@{}}
\toprule
&
\multicolumn{2}{c}{\textbf{Task performance}}
&
\multicolumn{2}{c}{\textbf{Per instance}}
&
\multicolumn{2}{c}{\textbf{Optimizer overhead}} \\
\cmidrule(lr){2-3}
\cmidrule(lr){4-5}
\cmidrule(l){6-7}

\textbf{Method}
    & \textbf{Score} $\uparrow$
    & \makecell{\textbf{$\Delta$}\\(pp)}
    & \makecell{\textbf{Cost}\\(\$) $\downarrow$}
    & \makecell{\textbf{Latency}\\(s) $\downarrow$}
    & \makecell{\textbf{Cost}\\(\$) $\downarrow$}
    & \makecell{\textbf{Wall-clock}\\(h:mm:ss) $\downarrow$} \\
\midrule

\textbf{AppWorld} & \emph{TGC (\%)} \\
\addlinespace[2pt]

\textbf{\textsc{DeCAF} (full)}
    & 84.23\se{0.89}
    & 0.00
    & 0.1840
    & 32.81
    & 191.86
    & 04:27:57 \\

\quad No Dimensions
    & 79.17\se{1.19}
    & --5.06
    & 0.2273
    & 56.63
    & 153.59
    & 02:59:56 \\

\quad Single Shared Session
    & 80.67\se{0.84}
    & --3.56
    & 0.2349
    & 55.92
    & 174.80
    & 07:01:55 \\

\quad No Experiment Proposals
    & 74.40\se{2.98}
    & --9.83
    & 0.3838
    & 78.47
    & 134.23
    & 06:12:08 \\

\quad No \texttt{node\_eval}
    & 83.93\se{0.60}
    & --0.30
    & 0.1971
    & 41.61
    & 181.42
    & 05:15:36 \\

\quad No Synthesis
    & 84.18\se{0.72}
    & --0.05
    & 0.1840
    & 32.82
    & 178.05
    & 04:07:23 \\

\quad Dimension Best
    & 87.20\se{0.89}
    & +2.97
    & 0.4273
    & 114.34
    & 175.48
    & 04:00:57 \\

\midrule

\textbf{LiveCodeBench} & \emph{Pass@1 (\%)} \\
\addlinespace[2pt]

\textbf{\textsc{DeCAF} (full)}
    & 53.16\se{0.73}
    & 0.00
    & 0.0068
    & 27.9
    & 81.88
    & 04:36:47 \\

\quad No Dimensions
    & 49.93\se{1.12}
    & --3.23
    & 0.0065
    & 34.5
    & 80.59
    & 04:10:01 \\

\quad Single Shared Session
    & 52.89\se{0.93}
    & --0.27
    & 0.0068
    & 28.4
    & 83.01
    & 07:49:36 \\

\quad No Experiment Proposals
    & 49.52\se{0.87}
    & --3.64
    & 0.0069
    & 30.2
    & 74.61
    & 04:40:29 \\

\quad No \texttt{node\_eval}
    & 50.10\se{0.68}
    & --3.06
    & 0.0070
    & 29.0
    & 76.02
    & 04:32:57 \\

\quad No Synthesis
    & 44.81\se{0.84}
    & --8.35
    & 0.0072
    & 31.1
    & 72.72
    & 04:02:11 \\

\quad Dimension Best
    & 51.37\se{0.90}
    & --1.79
    & 0.0068
    & 29.7
    & 71.15
    & 03:55:48 \\

\bottomrule
\end{tabular}
\end{table*}

\textbf{Node-level diagnosis.}\quad
The ablations establish that \texttt{node\_eval} improves final synthesis, but do not determine whether its component-level diagnoses accurately identify the source of a failure. We therefore sample 100 execution trajectories and inject a known failure into a randomly selected component of each trajectory. We then run \texttt{node\_eval} on the corrupted trajectories and measure whether it identifies the injected component as the most likely source of failure or not. We measure exact top-1 localization against a random-component baseline of 31.8\%. \texttt{node\_eval} correctly localizes 77\% of injected failures. Manual inspection reveals two error cases: (1) When the immediate downstream component produces a lower-quality response due to the corrupted component, and is flagged as the source of failure and (2) over-reliance on the model-generated compliance rubric measuring whether the component follows its own system prompt. In such cases, \texttt{node\_eval} identifies the weakest observed output rather than the causal origin of the failure. Full details in Appendix~\ref{app:node-level}.

\section{Conclusion}

We presented \textsc{DeCAF}, a framework that turns agentic-pipeline or harness optimization from a monolithic search over a fixed subspace into an adaptive, decomposed, evidence-generating process. It starts by dividing the search space into benchmark and pipeline specific dimensions and then proposes diverse experiments for each dimension. It then optimizes each dimension with a coding-agent and finally composes all dimension-level findings into a final pipeline using a node-level trajectory analysis tool. This decomposed search allows \textsc{DeCAF} to achieve better performance--cost--latency frontier on four agentic benchmarks.

\section*{AI disclosure section}

In this work, we used generative AI tools for paper writing and coding our experiments. Our method and baselines themselves require LLMs as part of their methods. We have \textbf{not} used generative AI tools for interpreting results, refining hypothesis, or creating scientific images. We have reviewed all AI-assisted work. We thoroughly verify all AI generated artifacts. We take responsibility for the final content of this work.

\section*{Reproducibility Statement}

To ensure the reproducibility of our results, we release the algorithm of \textsc{DeCAF} in Appendix~\ref{app:algorithm} and prompts in the Appendix~\ref{app:prompts}. We also note that LLMs are inherently probabilistic in nature, and some results may vary upon each run.

\section*{Ethics Statement}

In accordance with ICLR policy, we disclose that AI assistants, specifically Grammarly for grammar correction and ChatGPT for sentence restructuring and paraphrasing, were utilized during the preparation of this manuscript. The authors have reviewed, edited, and take full responsibility for all final content presented in this paper.

\bibliography{iclr2027_conference}
\bibliographystyle{iclr2027_conference}

\clearpage
\appendix

\section{Detailed Results}
\label{app:full-results}

In this section, we describe the optimization steps done by \textsc{DeCAF} for all benchmarks along with the intermediate findings from each dimension. This discussion sheds more light to the main results present in Table~\ref{tab:main_results}.

\subsection{AppWorld}
\label{app:appworld-details}

Table~\ref{tab:appworld_results} shows the detailed results for the AppWorld benchmark. \textsc{DeCAF} optimized the seed pipeline of Sonnet-4.6-ReAct agent across four dimensions: prompts, agent configuration, pipeline topology, and context handling. Each dimension was given 6 experiments by the experiment generator.

The dimension prompt was tasked with optimizing the system prompt of the pipeline. While optimizing this dimension, the optimizer tried re-emphasizing reading API docs, few-shot examples, structured reasoning, changing answer format, adding common mistakes, and explicit instructions to focus on identity linking for APIs. The optimizer tried the first four out of these six suggested experiments and then conducted a final experiment based on the findings. It reasoned that best gains come when explicit API endpoints are listed in the prompt which reduced confusion. For example, the model often confused the API endpoint \texttt{show\_liked\_songs} and \texttt{show\_song\_library}. Along with this, it gave a few-shot example and final answer formatting.

The agent configuration dimension was tasked with optimizing the model and execution budgets. The optimizer tested increasing and decreasing the observation limit, changing the reasoning effort, extending the turn budget, and replacing backbone model (Sonnet-4.6) with Opus-4.8. Its main finding was that allowing longer observations reduced failures caused by incomplete API responses. Increasing reasoning effort or switching to Opus-4.8 did not resolve the failures observed under the original observation limit, while extending the turn budget provided little benefit because the inspected tasks finished before reaching it. Lower reasoning effort and shorter observations introduced additional interaction costs without improving task completion. The optimizer therefore retained the larger observation limit while preserving the original model and reasoning settings.

The pipeline topology dimension was tasked with changing the pipeline by introducing additional stages around the ReAct agent. The optimizer tried documentation priming, error recovery, post-completion verification, task planning, and answer grounding. Documentation priming was retained: a deterministic stage retrieved app and API descriptions before the ReAct loop and supplied them as reference material, reducing the agent’s need to discover this information through separate reasoning turns. Error recovery and answer grounding added overhead without improving completion. Verification and planning initially appeared promising on the small validation probe but regressed on the full validation set. The optimizer skipped the proposed history-compression stage because the inspected trajectories did not establish a need for it.

The context-handling dimension was tasked with optimizing how observations and interaction history were presented to the agent. The optimizer tried explicit truncation warnings, compression of older history, a persistent cache for retrieved API documentation, and dynamically expanding the observation limit. It retained the truncation warning, which explicitly identified incomplete output and instructed the agent to paginate or narrow its query before computing totals. Heuristic history compression lost details needed in subsequent turns, while the documentation-cache variants failed to prevent repeated lookups and increased interaction costs. Dynamic observation expansion appeared neutral on the small probe but regressed on full validation. The optimizer skipped removing successful code blocks and adding a running state summary because these proposals reused compression and relocation mechanisms that had already performed poorly.

The final synthesis stage combined the retained changes from all four dimensions: endpoint-selection guidance, the worked example and answer-formatting rules, the larger observation limit, documentation priming, and explicit truncation warnings.

\begin{table}[t]
\centering
\caption{
Comparison of optimization methods on the AppWorld test set. Task goal completion (TGC) and per-difficulty performance are reported in \% (mean\,$\pm$\,standard error). Per-instance columns report test-time cost (USD) and latency per task; optimizer columns report the total cost (USD) and wall-clock latency (h:mm:ss) of the optimizer. $\uparrow$/$\downarrow$ indicate whether higher or lower is better.
}
\label{tab:appworld_results}
\fontsize{8pt}{9.5pt}\selectfont
\setlength{\tabcolsep}{3pt}
\begin{tabular}{@{}l c ccc cc cc@{}}
\toprule
 & & \multicolumn{3}{c}{\textbf{Test difficulty}\,$\uparrow$}
   & \multicolumn{2}{c}{\textbf{Per instance}\,$\downarrow$}
   & \multicolumn{2}{c}{\textbf{Optimizer}\,$\downarrow$} \\
\cmidrule(lr){3-5} \cmidrule(lr){6-7} \cmidrule(l){8-9}
\textbf{Method} & \textbf{TGC}\,$\uparrow$ & \textbf{1} & \textbf{2} & \textbf{3}
   & \textbf{Cost} (\$) & \textbf{Latency} (s) & \textbf{Cost} (\$) & \textbf{Latency} \\
\midrule
\emph{Base MAS} \\
\quad Opus-4.8-High     & 87.20\se{0.89} & 94.74\se{0.00} & 93.75\se{0.00} & 75.40\se{2.38} & 0.4273 & 114.3 & -- & -- \\
\quad Sonnet-4.6-Medium & 78.17\se{1.21} & 86.55\se{0.59} & 84.03\se{2.50} & 66.14\se{1.91} & 0.2242 & 55.1 & -- & -- \\
\quad Haiku-4.5-Low     & 35.71\se{0.59} & 64.91\se{5.27} & 31.25\se{0.00} & 12.70\se{6.35} & 0.1094 & 86.5 & -- & -- \\
\addlinespace
GEPA                    & 74.40\se{1.19} & 84.21\se{5.26} & 79.17\se{0.00} & 63.49\se{3.17} & 0.2901 & 58.8 & 279.04 & 14:56:44 \\
MASPO                   & 78.17\se{1.21} & 86.55\se{0.59} & 84.03\se{2.50} & 66.14\se{1.91} & 0.2242 & 55.1 & 103.57 & 02:05:37 \\
Meta-Harness            & 80.06\se{0.89} & 87.72\se{0.00} & 88.54\se{1.04} & 66.67\se{3.17} & 0.2237 & 73.8 & 166.91 & 07:58:13 \\
\midrule
\textbf{\textsc{DeCAF} (Ours)} & 84.23\se{0.89} & 90.35\se{0.88} & 89.58\se{0.00} & 74.60\se{1.59} & 0.1840 & 32.8 & 191.86 & 04:27:57 \\
\bottomrule
\end{tabular}
\end{table}

\subsection{LiveCodeBench}
\label{app:livecodebench-details}
\begin{table}[t]
\centering
\caption{
Comparison of optimization methods on the LiveCodeBench test set. Total and per-difficulty pass@1 are reported in \% (mean\,$\pm$\,standard error). Per-instance columns report test-time cost (USD) and latency per task; optimizer columns report the total cost (USD) and wall-clock latency (h:mm:ss) of the optimizer. $\uparrow$/$\downarrow$ indicate whether higher or
lower is better.
}
\label{tab:livecodebench_results}
\fontsize{8pt}{9.5pt}\selectfont
\setlength{\tabcolsep}{5pt}
\begin{tabular}{@{}l ccc cc cc@{}}
\toprule
 & \multicolumn{3}{c}{\textbf{pass@1}\,$\uparrow$}
   & \multicolumn{2}{c}{\textbf{Per instance}\,$\downarrow$}
   & \multicolumn{2}{c}{\textbf{Optimizer}\,$\downarrow$} \\
\cmidrule(lr){2-4} \cmidrule(lr){5-6} \cmidrule(l){7-8}
\textbf{Method} & \textbf{Total} & \textbf{Medium} & \textbf{Hard}
   & \textbf{Cost} (\$) & \textbf{Latency} (s) & \textbf{Cost} (\$) & \textbf{Latency} \\
\midrule
\emph{Base MAS} \\
\quad GPT-5.4-nano & 47.33\se{1.13} & 59.88\se{2.55} & 39.20\se{0.33} & 0.0051 & 29.3 & -- & -- \\
\quad GPT-5.4-mini & 44.50\se{1.78} & 65.84\se{2.29} & 30.67\se{1.49} & 0.0085 & 10.5 & -- & -- \\
\quad GPT-5.4      & 55.34\se{0.56} & 82.31\se{1.09} & 37.87\se{0.27} & 0.0236 & 12.5 & -- & -- \\
\addlinespace
GEPA & 48.30\se{1.70} & 62.96\se{3.70} & 38.80\se{0.40} & 0.0055 & 37.2 & 24.60 & 9:30:41 \\
MASPO               & 48.30\se{0.24} & 61.11\se{0.62} & 40.00\se{0.80} & 0.0052 & 26.4 & 74.21 & 2:48:15 \\
Meta-Harness        & 51.70\se{1.21} & 65.43\se{0.00} & 42.80\se{2.00} & 0.0065 & 44.4 & 57.22 & 7:04:25 \\
\midrule
\textbf{\textsc{DeCAF} (Ours)} & 53.16\se{0.73} & 75.31\se{0.00} & 38.80\se{1.20} & 0.0068 & 27.9 & 81.88 & 4:36:47 \\
\bottomrule
\end{tabular}
\end{table}

Table~\ref{tab:livecodebench_results} shows the detailed results of all optimizers on LiveCodeBench. \textsc{DeCAF} optimized a sequential pipeline consisting of an initial predictor, a reflector, and a final predictor, with all stages initially using GPT-5.4-nano. The pipeline generated a draft solution, critiqued it, and produced a revised solution for evaluation. Optimization explored prompts, agent configuration, pipeline topology, and context handling, with six experiments for each dimension.

The prompt dimension was tasked with optimizing the system prompt instructions given to these three agents. The optimizer tried explicit algorithm-selection reasoning, an edge-case checklist for the reflector, adversarial reimplementation instructions for the final predictor, and prompted self-verification. These changes reduced performance on medium-difficulty problems without producing consistent improvements on hard problems. The optimizer skipped the proposed complexity-reasoning preamble and few-shot reasoning example after judging them similar to previous unsuccessful experiments. An additional experiment suggested by the optimizer introduced a minimal instruction to preserve code that the critique had not flagged, but this also failed to improve overall completion. Consequently, no changes from the prompt dimension were retained.

The agent configuration dimension was tasked with optimizing model assignments and token budgets. The optimizer tested replacing GPT-5.4-nano with GPT-5.4-mini in each stage individually, in both predictors together, and throughout the pipeline. None of these configurations produced an accepted improvement. Upgrading the reflector redistributed successes between difficulty levels without improving overall completion, while upgrading both predictors and retaining the nano reflector produced a particularly unfavorable combination. Trace inspection also identified algorithmically incorrect solutions despite well-formed, untruncated output. The proposed token-budget reallocation was skipped because the inspected generations remained below the existing limits. The optimizer therefore retained the original configuration.

The pipeline topology dimension was tasked with exploring additional stages and alternative pipeline configurations. The optimizer tried candidate sampling and voting, algorithm planning, debate between independent solutions, confidence-gated retries, synthetic test-case generation, and complexity analysis. The retained change was a complexity-analysis stage that extracted input bounds, derived suitable time and space complexity requirements, and suggested compatible algorithm families before drafting began. This supplied the initial predictor with a concise constraint on its solution strategy while leaving implementation details open. The pipeline involving the planning stage regressed, and the confidence gate triggered retries, providing little selectivity. Sampling, debate, and synthetic tests also failed to justify their additional cost and latency. The final topology placed complexity analysis before the original predictor–reflector–predictor chain.

The context-handling dimension was tasked with improving how problems, code, and critiques were presented between stages. The optimizer tried XML-tagged messages, explicit input/output headers, problem-derived failure checklists, annotated code, parsed example tables, and structured critique fields. It retained a failure checklist generated from problem keywords and constraints, which directed the reflector toward potential issues such as indexing errors, graph traversal mistakes, and incorrect dynamic-programming base cases. This change showed a modest benefit when evaluated independently. The other representations did not improve overall completion, and splitting critiques into separate fields regressed. Combining the checklist with the parsed example table also performed worse than either change alone, so the dimension retained only the checklist and its supporting reflector instruction.

The final synthesis stage combined the complexity-analysis stage with the reflector’s failure checklist, the two changes retained by the individual dimensions. Although each had helped independently, their combination regressed in validation. Component-level diagnostic tool suggested that complexity analysis changed the initial predictor’s algorithmic approach, while the checklist continued to propose checks based only on keywords in the original problem. Requiring the reflector to address every checklist item could therefore direct its critique toward issues that were poorly matched to the chosen solution, prompting harmful revisions. The optimizer removed the checklist and retained the complexity-analysis stage, together with the supporting prompt and context changes needed to convey its output to the initial predictor. The final pipeline performed complexity analysis, draft generation, critique, and refinement, with all stages using GPT-5.4-nano.

\subsection{AgentDojo}
\label{app:agentdojo-details}

\begin{table}[t]
\caption{
Results on AgentDojo. Utility and attack success rate (ASR) are reported in \%, with the run-level standard error as subscript. $\downarrow$: lower is better.
}
\label{tab:agentdojo_results}
\begin{center}
\footnotesize
\setlength{\tabcolsep}{2pt}
\begin{tabular}{lcccccccc}
\toprule
 & \multicolumn{3}{c}{\textbf{Performance (\%) ($\uparrow$)}} & & \multicolumn{2}{c}{\textbf{Per instance}} & \multicolumn{2}{c}{\textbf{Optimizer}} \\
\cmidrule(lr){2-4} \cmidrule(lr){6-7} \cmidrule(lr){8-9}
\textbf{Method} & \textbf{\shortstack{Under\\attack}} & \textbf{Benign} & \textbf{Attacked} & \textbf{\shortstack{ASR\\($\downarrow$)}} & \textbf{Cost (\$)} & \textbf{\shortstack{Latency\\(s)}} & \textbf{Cost (\$)} & \textbf{\shortstack{Latency}} \\
\midrule
\multicolumn{9}{l}{\textit{Base MAS}} \\
\hspace{0.5em} Qwen-3.5-397b-a17b-FP8 & 38.6\se{1.6} & 96.1\se{1.3} & 65.3\se{0.3} & 55.5\se{1.0} & 0.0163 & 16.5 & -- & -- \\
\hspace{0.5em} Gemma-4-31B-it         & 69.5\se{0.7} & 90.3\se{0.7} & 82.8\se{0.3} & 18.8\se{0.7} & 0.0060 & 49.3 & -- & -- \\
\addlinespace
GEPA         & 82.1\se{1.0} & 86.4\se{0.7} & 82.8\se{1.0} & 0.7\se{0.0} & 0.0137 & 24.2 & 32.92 & 4:17:37 \\
MASPO        & 85.1\se{1.3} & 92.2\se{0.0} & 86.0\se{1.0} & 1.6\se{0.3} & 0.0147 & 36.7 & 21.05 & 1:14:38 \\
Meta-Harness & 85.4\se{0.3} & 89.0\se{0.7} & 85.7\se{0.0} & 1.0\se{0.3} & 0.0133 & 37.4 & 28.74 & 7:11:48 \\
\midrule
\textbf{\textsc{DeCAF} (Ours)} & 85.4\se{1.0} & 91.6\se{0.7} & 86.4\se{0.7} & 0.7\se{0.0} & 0.0126 & 25.6 & 26.48 & 4:43:13 \\
\bottomrule
\end{tabular}
\end{center}
\end{table}

Table~\ref{tab:agentdojo_results} shows the detailed results for the AgentDojo benchmark. \textit{Under Attack} represents success when attacked failed. \textit{Benign} represents clean success. \textit{Attacked} represents success in the presence of an attack, regardless of whether the attack succeeded or not. \textit{ASR} represents how often the attack actually succeeds. \textsc{DeCAF} optimized a single-stage Qwen-3.5 ReAct agent across four dimensions: prompts, agent configuration, pipeline topology, and context handling. Each dimension received six proposed experiments. Optimization considered benign task utility, utility under the evaluated prompt-injection attack, attack success rate, cost and latency as objectives.

The prompt dimension explored explicit trust boundaries, user-intent grounding for state-changing actions, deliberation over suspicious instructions, and source-attributed action justifications. The most effective retained change explicitly distinguished user instructions from tool-returned data, directing the agent to ignore embedded instructions even when they claimed administrative authority, while continuing the legitimate task. Adding user-intent grounding further improved utility under attack but reduced benign utility and increased cost, so it remained an alternative rather than replacing the simpler prompt. Preliminary probes of additional deliberation and source-attributed justifications showed no improvement in utility or security and added cost. Pre-action intent locks and instruction-source labels were proposed but skipped because they overlapped with the defenses already explored.

The agent-configuration dimension explored model selection, output-token budgets, iteration limits, decoding temperature, and adaptive token allocation. Replacing with Gemma-4-31B reduced cost and attack success but sacrificed benign utility, while tightening Qwen’s output budget provided another cost–utility trade-off. The optimizer also found that iteration limits interacted with model choice: reducing the iteration budget had little effect on the original Qwen configuration, whereas a tighter limit substantially reduced Gemma’s cost at some expense to robustness under attack. The temperature experiment yielded no significant difference, and adaptive per-turn token allocation required implementation changes outside this dimension’s scope. Consequently, the dimension retained several alternatives representing different utility, security, and cost trade-offs.

The pipeline-topology dimension tested a pre-execution intent-anchor stage and an intermediate tool-output sanitization step. The intent anchor generated a contract describing the legitimate task and permitted actions, while the sanitizer used an additional model call to remove instruction-like content from tool observations. Both improved attack resistance relative to the seed, but their additional cost and utility trade-offs made them dominated by the simpler prompt-based defense. Proposed post-execution auditing and final ratification were skipped because their read-only checks could not alter actions that had already occurred. Dual-track verification and hierarchical control were also skipped because the executor did not expose the intermediate reasoning and control hooks. No topology change was retained.

The context-handling dimension explored recency-based windowing, salience-based pruning, context resets, persistent user-intent anchors, source tagging, and reverse-recency ordering. The recency-window experiment worsened attack-related outcomes and increased cost. However, its implementation also replaced native tool-call messages with manually reconstructed context, so the regression could not be attributed to windowing alone. Source tagging was more promising: explicitly marking tool results as untrusted data preserved benign utility and improved attack resistance while retaining the native execution loop. The remaining proposals were skipped because they required context-control mechanisms unavailable in the frozen executor or similar reconstruction of the conversation history. Source tagging was retained within the dimension, but its candidate was dominated by the prompt-based alternatives in the shared archive.

The final synthesis phase combined the security-boundary prompt with the model and budget changes. Its representative pipeline retained the single-stage Qwen agent with the strengthened prompt and a tighter output budget, reducing cost relative to the prompt-only candidate while sacrificing some utility under attack. Relaxing the output budget increased cost without improving utility or security. Applying the same prompt to Gemma produced additional lower-cost alternatives, and subsequent iteration and output limits exposed further trade-offs. The optimizer also revealed an interaction between budgets: Qwen’s iteration limit became consequential after shorter outputs increased the number of turns needed to complete tasks. Additional topology stages and source tagging were not combined with the prompt, since their standalone results provided no clear justification for doing so. The final archive therefore retained multiple nondominated configurations rather than a single configuration that was best on every objective. We report the test set performance on the pipeline the optimizer chose at the end.

\subsection{PostTrainBench}
\label{app:posttrainbench}

\begin{table}[t]
\centering
\caption{Per-model breakdown of the OOD test results of PostTrainBench (mean\,$\pm$\,standard error). Each method is evaluated on two held-out base models and three benchmarks. Their average and the cheating rate are reported in \%. Optimizer columns report total cost (USD) and wall-clock latency (h:mm:ss); per-instance columns report the averaged cost and latency over the test instances. $\uparrow$/$\downarrow$ indicate whether higher or lower is better.}
\label{tab:posttrainbench_detailed}
\fontsize{8pt}{9pt}\selectfont
\setlength{\tabcolsep}{3pt}
\begin{tabular*}{\textwidth}{@{\extracolsep{\fill}}l ccccc rr rr@{}}
\toprule
 & \multicolumn{5}{c}{\textbf{OOD test}\,$\uparrow$}
   & \multicolumn{2}{c}{\textbf{Per instance}\,$\downarrow$}
   & \multicolumn{2}{c}{\textbf{Optimizer}\,$\downarrow$} \\
\cmidrule(lr){2-6} \cmidrule(lr){7-8} \cmidrule(l){9-10}
\textbf{Method} & \textbf{HumanEval} & \textbf{AIME 2026} & \textbf{BFCL} & \textbf{Average}
   & \textbf{Cheat.}\,$\downarrow$
   & \textbf{Cost} & \textbf{Latency}
   & \textbf{Cost} & \textbf{Latency} \\
\midrule

\emph{Base model} \\
\quad Qwen3-4B   & 31.71\se{0.00} & 1.67\se{0.96} & 38.16\se{0.01} & 23.85\se{0.32} & -- & -- & -- & -- & -- \\
\quad Gemma-3-4B & 0.61\se{0.00}  & 0.00\se{0.00} & 19.73\se{0.00} & 6.78\se{0.00}  & -- & -- & -- & -- & -- \\
\addlinespace

\emph{Instruct model} \\
\quad Qwen3-4B   & 76.22\se{0.00} & 52.50\se{2.10} & 62.73\se{0.07} & 63.82\se{0.70} & -- & -- & -- & -- & -- \\
\quad Gemma-3-4B & 71.49\se{0.46} & 9.17\se{2.50}  & 39.32\se{0.05} & 39.99\se{0.85} & -- & -- & -- & -- & -- \\
\addlinespace

\emph{Claude Code} \\
\emph{(Sonnet-5-High)}\\
\quad Qwen3-4B   & 49.09\se{2.74} & 10.00\se{3.33} & 52.33\se{2.21} & 37.14\se{1.62} & 25.00 & 10.98 & 08:11:57 & -- & -- \\
\quad Gemma-3-4B & 36.59\se{7.93} & 0.00\se{0.00}  & 53.36\se{0.71} & 29.98\se{2.65} & 0.00  & 9.15 & 07:49:21 & -- & -- \\
\addlinespace

\emph{Claude Code} \\
\emph{(Opus-4.8-Max)}\\
\quad Qwen3-4B   & 78.05\se{3.58} & 0.80\se{0.80} & 56.93\se{2.73} & 45.26\se{1.52} & 0.00 & 21.21 & 08:21:27 & -- & -- \\
\quad Gemma-3-4B & 53.82\se{2.59} & 0.00\se{0.00} & 53.61\se{1.86} & 35.81\se{1.06} & 0.00 & 19.18 & 08:12:06 & -- & -- \\
\addlinespace

\emph{GEPA} \\
\quad Qwen3-4B   & 49.09\se{2.74} & 10.00\se{3.33} & 52.33\se{2.21} & 37.14\se{1.62} & 25.00 & 10.98 & 08:11:57 & 154.26 & 20:40:51 \\
\quad Gemma-3-4B & 36.59\se{7.93} & 0.00\se{0.00}  & 53.36\se{0.71} & 29.98\se{2.65} & 0.00  & 9.15  & 07:49:21 & 154.26 & 20:40:51 \\
\addlinespace

\emph{MASPO} \\
\quad Qwen3-4B   & 49.09\se{2.74} & 10.00\se{3.33} & 52.33\se{2.21} & 37.14\se{1.62} & 25.00 & 10.98 & 08:11:57 & 86.10 & 6:20:49 \\
\quad Gemma-3-4B & 36.59\se{7.93} & 0.00\se{0.00}  & 53.36\se{0.71} & 29.98\se{2.65} & 0.00  & 9.15  & 07:49:21 & 86.10 & 6:20:49 \\
\addlinespace

\emph{Meta-Harness} \\
\quad Qwen3-4B   & 71.65\se{7.02} & 3.33\se{3.33} & 32.43\se{12.67} & 35.80\se{4.95} & 12.50 & 11.04 & 08:25:46 & 145.63 & 42:45:43 \\
\quad Gemma-3-4B & 36.89\se{2.47} & 3.33\se{0.00} & 44.16\se{6.98} & 28.13\se{2.47} & 12.50 & 8.75 & 05:41:41 & 145.63 & 42:45:43 \\
\addlinespace

\textsc{DeCAF} \\
\quad Qwen3-4B   & 64.63\se{4.27} & 3.33\se{3.33} & 58.33\se{0.11} & 42.10\se{1.81} & 0.00 & 11.98 & 04:31:09 & 177.25 & 43:51:30 \\
\quad Gemma-3-4B & 30.47\se{2.34} & 0.00\se{0.00} & 56.75\se{0.15} & 29.07\se{0.83} & 0.00 & 8.06 & 03:09:57 & 177.25 & 43:51:30 \\
\bottomrule
\end{tabular*}
\end{table}

Table~\ref{tab:posttrainbench_detailed} shows the detailed results for the PostTrainBench benchmark. \textsc{DeCAF} optimized a seed pipeline in which a single Sonnet-5-High Claude Code agent was responsible for training a given model for a given benchmark. It started by proposing seven dimensions: capability and budget allocation, checkpoint probing and selection, domain-adaptive data curation, multi-role topology with supervision, publication integrity and fallback, structured context handoffs, and training-method design. The experiment generator proposed two experiments for each dimension. As per the PostTrainBench benchmark design, each pipeline is given a total of 10hrs to complete their training.

The capability and budget allocation dimension was tasked with deciding how agent capability and the wall-clock budget should be divided across the pipeline. The optimizer first separated the work into a low-effort scouting stage and a high-effort builder stage, and later revised this design so that a usable checkpoint was published before optional evaluation. It also tested a throughput-aware cutoff that was intended to stop training early enough to preserve time for checkpoint publication. The scout was inexpensive and often created usable artifacts quickly, while publishing before evaluation protected the result from late failures. However, none of the standalone variants improved the validation objective: one failed to produce a compatible checkpoint, a longer run reached its 10hr deadline, and the adaptive cutoff was never needed because the agents finished naturally. The final pipeline therefore did not retain the additional scout role or a second scheduling policy. It retained only the general lesson of reserving time for publication and publishing a usable checkpoint before the 10hr time limit.

The checkpoint probing and selection dimension was tasked with choosing the best locally produced checkpoint. The optimizer tested a deterministic held-out negative-log-likelihood selector over training data and a synthetic selector based on output-format validity and self-consistency. Both mechanisms produced clean, loadable checkpoints and performed selection. Nevertheless, neither correlated reliably with downstream validation performance, and reserving time for probing reduced the time available for training. Both selectors were therefore removed from the final pipeline.

The domain-adaptive data curation dimension was tasked with selecting training data using only the immutable task description and other benchmark-safe information. The first experiment classified the task into broad signatures, such as procedural mathematics, verifiable reasoning, or broad knowledge, and used a bounded curator to assemble a compact corpus together with provenance and license metadata. This approach improved both validation probes at low cost and was retained. The second experiment added deterministic structural and executable filters, including arithmetic checks, valid choice-label checks, completeness checks, and deduplication. Although it helped on one probe, it regressed on the other and consumed substantially more of the budget. The optimizer therefore retained coarse domain routing and provenance validation but dropped aggressive task-specific filtering.

The multi-role topology with supervision dimension was tasked with replacing the single-agent pipeline with specialized agents and explicit artifact gates. One experiment used four sequential roles for design, training, local selection, and delivery, with a supervisor responsible for retrying failed stages. A second experiment created two independently planned and trained strategies, divided the training budget between them, and selected between their checkpoints using training-local likelihood. The four-role topology failed the integrity check (cheating detected by the LLM judge) and was disqualified despite a promising raw result, while the competing-strategy topology diluted a short training budget across two branches and did not improve performance. The additional roles also introduced coordination overhead, and one branch spent time searching for timing and environment information rather than training. The final pipeline therefore used a small number of roles with deterministic handoffs instead of a supervisor or competing training branches.

The publication integrity and fallback dimension was tasked with ensuring that a valid model survived timeouts, crashes, and incomplete final saves. The optimizer tested a watchdog that validated snapshots and atomically published the newest usable checkpoint, with the base model as a final fallback. It also tested measuring checkpoint-save latency and deriving a hard training stop that left enough time for a final save and publication. After several implementation failures in the experiment harness were repaired, both mechanisms produced clean, loadable checkpoints. Neither improved the validation objective, however, and their intended crash-recovery paths were not exercised because the training agents normally finished on their own. The final pipeline omitted the watchdog and duplicate deadline controller, but retained a single publication owner, atomic publication, and a protected end-of-run reserve.

The structured context handoff dimension was tasked with transferring decisions between pipeline stages without carrying a long conversational history. The optimizer tested a fixed-schema JSON handoff containing the selected data, hyperparameters, elapsed budget, target metric, checkpoint path, and completion status. It also tested a capped rolling ledger containing recent metrics, their slope, plateau status, and the best checkpoint. The compact fixed-schema handoff transferred the important decisions more effectively, whereas the rolling metric ledger added context without improving the result and regressed on validation. The final pipeline therefore retained only compact, deterministically validated manifests and receipts. It did not retain a rolling ledger.

The training-method design dimension was tasked with choosing both the optimization method and its schedule under the available data and time budget. The first experiment introduced a verifiability-gated policy: tasks with a reliable local checker could use rejection-filtered supervised fine-tuning, while other tasks used direct supervised fine-tuning; preference-style refinement was optional when suitable comparisons could be formed safely. The second experiment added a short real training probe, measured optimizer-step throughput, and deterministically derived the total step count and checkpoint cadence from the remaining deadline. The measured schedule performed better overall. Unlike several proposed adaptive mechanisms in the other dimensions, this policy demonstrably changed the executed training schedule. The final pipeline retained the method classifier and the empirical throughput probe, with this stage serving as the sole owner of the training-step and checkpoint schedule.

The final synthesis combined the mechanisms from several dimensions. A bounded, non-GPU curation stage selected a compact and legally documented corpus using a coarse domain signature. A deterministic manifest then handed the result to one training stage, which selected direct or rejection-filtered supervised fine-tuning based on whether a trustworthy local verifier was available. A short micro-training run measured actual throughput, after which the remaining steps and checkpoint cadence were computed from the deadline while preserving a publication reserve. Deterministic gates checked the handoff and final artifact, and atomically published the selected checkpoint. The synthesis optimizer excluded evaluator feedback, benchmark examples, hosted teacher labels, checkpoint-quality proxies, competing training branches, rolling metric histories, a watchdog, and redundant deadline controllers.

\section{Node-Level Diagnostics}
\label{app:node-level}

\begin{table}[t]
\centering\small
\caption{Controlled failures used to evaluate \texttt{node\_eval}. Corruptions preserve the format and style of the original output, contain no marker of the injected location, and in several cases produce a visible defect only in a downstream component.}
\begin{tabular}{p{0.20\linewidth}p{0.33\linewidth}p{0.37\linewidth}}
\toprule
Failure type & Injection & Example \\
\midrule
Incorrect contribution & Modify a task-relevant conclusion or intermediate result while preserving the surrounding output. & LCB reflector: a critique that flags a correct line as a bug and proposes a concrete fix; the refiner applies it, so the defect surfaces in the refiner's output. \\
Input unfaithfulness & Alter or contradict a constraint, observation, or fact supplied to the component. & AppWorld agent: the reasoning step summarizes a truncated API observation as complete, ignoring the truncation warning, and computes a total from the partial page. \\
Missing information & Remove information required by a downstream component or output contract. & LCB reflector: delete the critique item on which the final fix depends (identified by counterfactual re-execution), keep the remaining valid items, and renumber; the critique stays well-formed. \\
Instruction non-compliance & Violate one explicit requirement derived from the component's system prompt. & AgentDojo agent: its prompt requires ignoring instructions found in tool output; the agent executes one such instruction while still completing the user task. \\
\bottomrule
\end{tabular}
\label{tab:node_eval_faults}
\end{table}

The ablations in Section~\ref{sec:results-ablation} measure whether \texttt{node\_eval} helps the synthesis optimizer, but not whether its component-level diagnoses point at the right component. We therefore construct a controlled fault-injection evaluation in which the source of each failure is known by construction.

\paragraph{Trajectory sampling.}
We sample 100 execution trajectories with complete component-level traces from AppWorld, LiveCodeBench, and AgentDojo. To avoid ambiguity from pre-existing failures, we use only trajectories on which the pipeline succeeded. For each trajectory we sample one component uniformly and inject exactly one failure into its output. The sampled component, the failure class, and all corruption metadata are hidden from \texttt{node\_eval}.

\paragraph{Failure classes.}
We consider four classes of component failure, each targeting one aspect that \texttt{node\_eval} is designed to assess: (1)~\emph{incorrect contribution}, the component's substantive decision, computation, or conclusion is wrong; (2)~\emph{input unfaithfulness}, the component contradicts or incorrectly transforms information supplied to it; (3)~\emph{missing information}, the component omits task-relevant information it is responsible for producing or passing downstream; and (4)~\emph{instruction non-compliance}, the component violates an explicit requirement of its own system prompt. Table~\ref{tab:node_eval_faults} gives one example per class.

\paragraph{Corruption construction.}
Corruptions are designed to be plausible rather than trivially detectable. Each corruption is generated by prompting GPT-5.6-Sol with the clean output and a class-specific instruction, then reviewed by an author under three rules: (i)~the corrupted output preserves the original format, length, style, and all task-irrelevant content; (ii)~exactly one task-relevant aspect is changed, and content that depends on it is updated so the output stays internally consistent (for example, a misreported input bound is accompanied by a matching change to the recommended complexity); (iii)~the output contains no marker, comment, or stylistic break indicating the edit.

\paragraph{Counterfactual execution.}
After modifying the selected component, we hold the trajectory prefix fixed and re-execute every downstream component on the corrupted output, so that downstream behavior and the final result are consistent with the injected failure and the error propagates naturally through the pipeline. The corrupted trajectory is then given to \texttt{node\_eval} without the clean output, the injected component, or the failure class.

\paragraph{Localization metric.}
We run the same \texttt{node\_eval} procedure and component-ranking rule used by the synthesis stage. Top-1 localization accuracy is the fraction of corrupted trajectories for which the top-ranked component is exactly the injected one. Because pipelines contain different numbers of evaluable components, the random-component baseline is
\begin{equation}
\frac{1}{N}\sum_{i=1}^{N}\frac{1}{|V_i|},
\end{equation}
where $V_i$ is the set of eligible components in trajectory $i$.

\paragraph{Results.}

\begin{table}[t]
\centering
\small
\caption{
Top-1 failure-localization accuracy of \texttt{node\_eval} under controlled component corruptions. Overall shows the accuracy of \texttt{node\_eval}, while Random Component shows the accuracy if a random component is returned every turn.
}
\begin{tabular}{lrr}
\toprule
Injected failure & $N$ & Top-1 localization (\%) \\
\midrule
Incorrect contribution      & 22 & 90.91 \\
Input unfaithfulness        & 29 & 79.31 \\
Missing information         & 22 & 86.36 \\
Instruction non-compliance  & 27 & 55.56 \\
\midrule
Overall                     & 100 & 77.00 \\
Random Component            & 100 & 31.81 \\
\bottomrule
\end{tabular}
\label{tab:node_eval_fault_results}
\end{table}
Table~\ref{tab:node_eval_fault_results} reports top-1 accuracy by failure class. \texttt{node\_eval} localizes 77\% of injected failures against a random baseline of 31.81\%. Accuracy is highest for incorrect contribution (90.91\%) and missing information (86.36\%), classes in which the defect can be checked against the component's input or its downstream contract, and lowest for instruction non-compliance (55.56\%).

Manual inspection of the 23 misses reveals two error modes. First, when the immediate downstream consumer produces a visibly weaker output because of the corrupted input, \texttt{node\_eval} flags the consumer rather than the source. Second, the compliance rubric that \texttt{node\_eval} derives from each component's prompt is over-weighted, so a component with a cosmetic violation is ranked above the causally responsible one. In both modes \texttt{node\_eval} identifies the weakest observed output rather than the causal origin of the failure.

These results indicate that the trajectory-level signals used by \texttt{node\_eval} reliably identify the source of controlled component failures and, where they miss, do so in a predictable direction, typically one hop downstream. We emphasize that the experiment evaluates localization under synthetic interventions with known ground truth on successful trajectories; it does not establish that \texttt{node\_eval} provides exact causal attribution for naturally occurring failures, which may involve several interacting defects.

\section{DeCAF Algorithm}
\label{app:algorithm}

Algorithm~\ref{alg:decaf} describes how \textsc{DeCAF} works for a given seed pipeline, metrics, and a benchmark.

\begin{algorithm}[t]
\caption{\textsc{DeCAF}: Divide, Diagnose, and Compose}
\label{alg:decaf}
\begin{algorithmic}[1]
\REQUIRE task specification $T$; development set $\Ddev$; seed pipeline $P_0$; optimization budget
\STATE $\{D_k=(\mathcal{M}_k,\mathcal{C}_k,\rho_k)\}_{k=1}^{K} \leftarrow \textsc{PlanDimensions}(T, \Ddev, P_0)$ \COMMENT{few-shot guided (Sec.~\ref{sec:decaf-divide})}
\FOR{$k = 1, \dots, K$ \textbf{in parallel}}
  \STATE $\mathcal{H}_k \leftarrow \textsc{GenerateHypotheses}(D_k)$ \COMMENT{iterative; each materially different}
  \STATE $\mathcal{A}_k \leftarrow \{P_0\}$;\ $\mathcal{F}_k \leftarrow \emptyset$;\ launch a coding session on a fresh copy of $P_0$, restricted to $\mathcal{M}_k$
  \WHILE{experiments remain within the session budget}
    \STATE select or adapt $h \in \mathcal{H}_k$; implement candidate $P'$
    \STATE evaluate $\mathbf{J}(P')$ on $\Ddev$ \COMMENT{Eq.~\ref{eq:objective-vector}}
    \STATE $\mathcal{A}_k \leftarrow \operatorname{ND}(\mathcal{A}_k \cup \{P'\})$; append the experiment's record to $\mathcal{F}_k$ \COMMENT{\texttt{FINDINGS.md}}
  \ENDWHILE
\ENDFOR
\STATE launch synthesis session with $P_0$ and $\{(\mathcal{A}_k, \mathcal{F}_k)\}_{k=1}^{K}$ \COMMENT{full edit access; \texttt{node\_eval} available only here}
\WHILE{the synthesis budget remains}
  \STATE compose promising findings; re-evaluate $\mathbf{J}$ end-to-end on $\Ddev$
  \STATE on regression: invoke \texttt{node\_eval} and apply targeted cross-dimension revisions
  \STATE $\mathcal{A}^{\star} \leftarrow \operatorname{ND}(\mathcal{A}^{\star} \cup \{\text{composed candidates}\})$
\ENDWHILE
\RETURN $\mathcal{A}^{\star}$; one final pipeline is selected from the archive by the optimizer for final test-set evaluation
\end{algorithmic}
\end{algorithm}

\section{Prompts}
\label{app:prompts}

The following sub-sections illustrate all the prompts used by different components of \textsc{DeCAF}.

\subsection{Dimension Proposal}
\label{app:dimension_proposal_prompt}

\paragraph{System Prompt}
\begin{promptbox}
You are the decomposition designer for DeCAF, an optimizer of compound agent systems. You are given a benchmark card (what the benchmark is, what the seed system is), the seed system's actual files, a sample of real task instances, and worked examples of good decompositions on other benchmarks.\newline

Your job: decide how the work of optimizing this seed on this benchmark should be SPLIT into multiple parallel focus-areas ("slices"), each optimized independently by its own coding-agent session. You define the SEARCH SUBSPACES; you do not search them.\newline

SLICES ARE GENERAL SCOPES, NOT EXPERIMENTS. A slice names the FAMILY of changes its session may explore and where that family lives in the seed. E.g. "prompt modification: rewrite/restructure/extend the system prompt text" or "topology optimization: add/modify/delete/reorder stages". Do NOT prescribe specific changes: no particular prompt habits or content, no named stages or mechanisms to add, no parameter values to try, no fix-for-a-failure. This ban includes illustrative lists: a focus must contain no "(e.g. ...)" enumerations of specific stages, habits, mechanisms, or techniques, name the change-family and its editable surface, then stop. (Enumerating the knobs that DEFINE a family - model / effort / budget scalars for the config family - is fine; enumerating candidate USES of a family is not.)\newline

The two properties your decomposition must have:\newline
- COVERAGE: collectively the slices span the WHOLE editable surface. Never leave a surface family out SILENTLY: forced coverage is the point of the method. If you judge a family genuinely not worth a dedicated slice for THIS benchmark, argue that explicitly in benchmark\_analysis, including where that family's levers fold into other slices as supporting changes. An argued omission is acceptable; a silent one is not.\newline
- SEPARABILITY: slices own DISTINCT change-families so parallel sessions explore different regions of the design space. Slices are soft (each session may make supporting changes on other surfaces in service of its own family), so mild overlap is fine - but two slices owning the same family is wasted parallelism.\newline

Rationales cite the card / seed files / instances as EVIDENCE OF LEVERAGE - reasons this change-family plausibly matters on this benchmark (e.g. an untouched surface, an observed failure class this family could address). State the evidence; do not turn it into the experiment.\newline

Output EXACTLY one JSON object, no prose before or after, in this schema:\newline
\{\newline
  "benchmark\_analysis": "<2-4 sentences: where on this benchmark the leverage is and why; argue any surface family you leave without a dedicated slice>",\newline
  "slices": [\newline
    \{\newline
      "name": "<short\_snake\_case, unique - becomes a folder name>",\newline
      "focus": "<1-2 sentences: the GENERAL scope - the family of changes this slice explores and where it lives in the seed; no specific experiments>",\newline
      "primary\_axis": "<dimension being optimized>",\newline
      "rationale": "<why this family has leverage on this benchmark — evidence, not fixes>"\newline
    \}\newline
  ]\newline
\}
\end{promptbox}

\paragraph{User Prompt}
\begin{promptbox}
\{\{BENCHMARK\_CARD\}\}\newline

\#\# Seed system files (verbatim)\newline

--- \{\{SYSTEM\_TXT\_FILE\_PATH\}\} ---\newline
\{\{SEED\_SYSTEM\_PROMPT\}\}\newline

--- \{\{SYSTEM\_CONFIG\_YAML\_PATH\}\} ---\newline
\{\{SEED\_SYSTEM\_CONFIG\}\}\newline

--- \{\{AGENT\_CONFIG\_YAML\_PATH\}\} ---\newline
\{\{SEED\_AGENT\_CONFIG\}\}\newline

\#\# Sampled task instances\newline

\{\{SAMPLED\_VALIDATION\_INSTANCES\}\}\newline

\{\{FEW\_SHOT\_DECOMPOSITION\_EXAMPLES\}\}\newline

---\newline

Now produce the decomposition for THIS benchmark card and seed. Respond with ONLY the JSON object.
\end{promptbox}

\subsection{Experiment Proposal}
\label{app:exp-prompt}

\paragraph{System Prompt}
\begin{promptbox}
You are the experiment ideation engine for DeCAF, an optimizer of compound agent systems. You are given a benchmark card (benchmark + seed system + editable surfaces + constraints), one slice of the optimization (its main focus), and the list of experiment ideas already proposed for that slice.\newline

Propose EXACTLY ONE new experiment idea for this slice - or stop.\newline

Rules:\newline
- The idea must serve the slice's MAIN FOCUS, be implementable within the editable surfaces, and respect every hard constraint on the card (temperature pinned at 1.0, frozen files, output contract). Supporting changes on other surfaces are allowed when they serve the focus.\newline
- DIVERSITY IS THE POINT: the new idea must rest on a genuinely DIFFERENT underlying concept or mechanism than EVERY previously proposed idea - a different lever, not a parameter variation, rewording, or combination of earlier ideas.\newline
- Propose only; do not implement, do not write code or full prompts; a concrete sketch of what to change and where is enough for the optimizer session to execute it.\newline
- Be concrete and falsifiable: say what would be edited, what effect is expected on the score vector (accuracy / cost / latency), and how a cheap probe would check it.\newline
- If no genuinely distinct, promising concept remains for this slice, STOP instead of padding with weak or duplicate ideas.\newline

Output EXACTLY one JSON object, no prose before or after. Either:\newline
\{\newline
  "title": "<short imperative name>",\newline
  "concept": "<one sentence: the underlying mechanism/lever this idea rests on>",\newline
  "change\_sketch": "<2-4 sentences: what to edit, where (files/surfaces), and roughly how>",\newline
  "expected\_effect": "<one sentence: which vector components should move and in which direction>",\newline
  "how\_to\_measure": "<one sentence: the cheap probe (--dev / --ids ...) and what confirms it>",\newline
  "est\_cost": "low | medium | high"\newline
\}\newline
or, to stop:\newline
\{ "stop": true, "reason": "<one sentence>" \}
\end{promptbox}

\paragraph{User Prompt}
\begin{promptbox}
\{\{BENCHMARK\_CARD\}\}\newline

\#\# Sampled task instances\newline

\{\{SAMPLED\_VALIDATION\_INSTANCES\}\}\newline

\#\# The slice you are ideating for\newline

- name: \{\{SLICE\_NAME\}\}\newline
- main focus: \{\{SLICE\_FOCUS\}\}\newline
- primary\_axis: \{\{PRIMARY\_AXIS\}\}\newline
- rationale: \{\{RATIONALE\}\}\newline

Previously proposed ideas for this slice - the new idea must use a DIFFERENT underlying concept than ALL of these:\newline
  E1. \{\{TITLE\}\} - concept: \{\{CONCEPT\}\} (sketch: \{\{CHANGE\_SKETCH\}\})\newline
  ...\newline
  E\{k-1\}. \{\{TITLE\}\} - concept: \{\{CONCEPT\}\} (sketch: \{\{CHANGE\_SKETCH\}\})\newline

Propose experiment idea \#\{\{k\}\} for this slice (or stop). Respond with ONLY the JSON object.
\end{promptbox}

\subsection{Dimension Coding-Agent}
\label{app:dimension_optimizer_prompt}

\paragraph{GOAL.md given as \texttt{/goal}}
\begin{promptbox}
\# Your goal - optimize the "\{\{NAME\}\}" slice\newline

Optimize this multi-agent system so it completes more tasks, with your **MAIN FOCUS on:\newline
\{\{FOCUS\}\}**  (primary axis: `\{\{PRIMARY\_AXIS\}\}`).\newline

Why this focus: \{\{RATIONALE\}\}\newline

Your main focus drives your work, but you MAY make supporting changes elsewhere in `pipeline/` when they genuinely serve that focus. What you must not do is drift into a *different* focus - that belongs to another slice.\newline

\#\# Objective\newline
Each finished task is graded by the benchmark's own objective reward. `report.py` prints the score vector:\newline
\{\{TASK\_METRIC\_INFORMATION\}\}\newline

\#\# Attempt 0 - the baseline you must beat\newline
The untouched seed in `../\_floor/`. Establish its vector (run your own `report.py --full` before any edit, or copy `../\_floor`'s row) as your Attempt 0. Your fork starts byte-identical to `../\_floor/pipeline/`.\newline

\#\# Search space - MAIN FOCUS: \{\{FOCUS\}\}\newline
\{\{DIMENSION\_DETAILS\}\}\newline

**Supporting changes (soft slice):** you may also touch the other dimensions above when a change there is *in service of your main focus* (e.g. a focus on "recovering from tool errors" may need both a prompt tweak and a retry in `agents.py`). Record them in `FINDINGS.md` as *supporting* the *primary* focus change.\newline

\#\# Experiment menu - \{\{MENU\_COUNT\}\} pre-proposed experiments (GuidedSampling)\newline

Before this session, an ideation pass proposed the experiments below for your focus. Each is built on a **different underlying concept**, listed in proposal order (earlier = higher expected value). The menu is a **map, not a mandate**:\newline

- **Work top-down.** Start from E1 and move down unless your own evidence redirects you; dip into the tail only when the head is exhausted or a trace points elsewhere.\newline
- **One attempt = one concept.** Implement one menu item per attempt, or one **off-menu** idea of your own (allowed and welcome; label it `off-menu` and say what prompted it). Never blend several menu concepts into a single attempt: per-concept attribution is the point, the merge phase reasons over which *concepts* worked.\newline
- **Log the menu id** (`E<n>` or `off-menu`) in every `JOURNAL.md` entry, and fill the **menu-outcome table in `FINDINGS.md`** (tried / skipped, verdict, one-line evidence) for every item before you stop - a skipped item with a reason is signal too.\newline
- A menu item is a starting hypothesis, not a spec - refine its concrete parameters from what the traces show, but keep the underlying concept recognizable.\newline

\{\{EXPERIMENT\_MENU\}\}\newline

\#\# Protocol - iterate cheap, accept on the full set\newline
- Probe with `report.py --dev` / `--ids <task\_id> ...`; accept only after `report.py --full`. Log every attempt to `JOURNAL.md` with the cost/latency it incurs.\newline
- Read the per-turn / per-stage traces in `runs/<ts>/<id>/` to see whether your change moved what you intended - and whether a supporting change earned its keep or was noise.\newline

\#\# Stopping - cover the menu's head, not one win\newline
Work through as many distinct menu concepts as your budget allows (at minimum, reach a verdict - tried or reasoned-skip - on the top half of the menu), weigh each against cost, then confirm your best on `--full` and **fill in `FINDINGS.md`**: the final config, the score vector vs Attempt 0 (cost/latency included), the **primary (focus) change vs supporting changes** split, the **menu-outcome table**, and the patch.
\end{promptbox}

\subsection{Final Synthesis Coding Agent}
\label{app:final_coding_prompt}

\paragraph{GOAL.md given as \texttt{/goal}}
\begin{promptbox}
\# Your goal - assemble the best whole-system agent\newline

You are given a multi-agent system plus several independent optimization results - each one improved the system along a different **focus**. **Assemble the best whole-system configuration** by combining them, guided by the per-component `node\_eval` diagnostic, and verify it composes. You have the FULL search space, seeded with those results. Operating rules are in `CLAUDE.md`.\newline

\#\# What you are handed\newline
Several independent optimization efforts, one per **focus** (their folders are listed in `../slices\_manifest.json`). Each optimized ONE main focus - but was allowed supporting changes on other axes - independently from the untouched baseline (`../\_floor/`), and left:\newline
- a `FINDINGS.md` (its **primary** focus change vs its **supporting** cross-axis changes, and its vector vs floor), and\newline
- a diverged `pipeline/`.\newline

Your fork also starts from `../\_floor/`, so every slice patch is a 3-way merge onto the same ancestor.\newline

\#\# Objective\newline
Same objective vector as the slices, graded by the benchmark's own objective reward \{\{REWARD INFORMATION\}\}\newline

\#\# The differentiator you have that the slices don't: node\_eval\newline
`diagnose.py` runs the frozen per-component `node\_eval` judge over a batch of runs and reports, per stage/component, **correctness** (advances a correct solution / faithful to input - primary) and **compliance** (obeys its own prompt - secondary) + the drift gap, weakest-correctness first. Use it to (a) rank the floor's weak components, and (b) catch composition regressions on candidate assemblies. It is **diagnostic only** - accept/reject is always the objective performance from `report.py`.\newline

\#\# Workflow\newline
1. Establish the floor vector (`../\_floor` `report.py --full`) and diagnose it.\newline
2. Read `../slices\_manifest.json`, then each slice's `FINDINGS.md`; apply each slice's **primary change with its supporting changes** onto your fork (3-way merge; reconcile hand-edits where two slices touched the same region). Because slices were soft, two slices may have changed the same file for different reasons - keep the change that serves the stronger effect, or combine them.\newline
3. Probe (`report.py --dev`/`--ids`), diagnose candidate assemblies (`diagnose.py runs/<ts>`), iterate\newline
- drop what regresses, keep what composes. Go **beyond** the slice patches where node\_eval points somewhere no slice explored.\newline
4. **End with a `--full` verification** of the final assembly.\newline

\#\# Stopping\newline
Stop when node\_eval shows no weak component you can still move and `--full` confirms your best assembly. Then **fill in `FINDINGS.md`**: what was taken from each slice, what was dropped and why, anything added beyond them, the final `--full` vector vs floor and vs each slice, and a single reproduce command.
\end{promptbox}

\subsection{node\_eval prompt}
\label{app:node_eval_prompt}

\paragraph{System Prompt}
\begin{promptbox}
You are a strict, evidence-grounded evaluator of ONE node in a
multi-agent pipeline, across SEVERAL execution instances.\newline

Your task is to assess:\newline
1. CORRECTNESS: whether the node produces a substantively sound contribution toward the user's goal.\newline
2. COMPLIANCE: whether the node follows its own system prompt.\newline
3. DOWNSTREAM UTILIZATION: whether its contributions are delivered, used, and retained in subsequent processing.\newline
4. TRAJECTORY: whether repeated iterations make meaningful progress.\newline

CORRECTNESS is primary. A node can comply perfectly with an inadequate instruction while producing an incorrect or unhelpful result. Conversely, a useful output can violate the node's specified format or procedure. Keep these judgments separate.\newline

Provide observational diagnostics, not claims of exact causal responsibility. Identify demonstrated defects, likely issue sources, and unresolved evidence gaps.\newline

Be demanding but fair. Do not invent flaws, penalize brevity, reward verbosity, or substitute skepticism for evidence.\newline

INPUTS AND TRUST BOUNDARIES\newline

Shared across instances:\newline
- node name;\newline
- the node's own system prompt, including its job description and output requirements.\newline

For each instance, the supplied record may contain:\newline
- instance and run identifiers;\newline
- user\_goal: the goal assigned to the complete pipeline;\newline
- the node's iteration-indexed inputs and outputs;\newline
- relevant tool actions, tool results, artifacts, or state changes;\newline
- downstream consumer identities and invocation identifiers;\newline
- the source iteration associated with each handoff;\newline
- consumer roles or input-output contracts;\newline
- actual inputs delivered to consumers;\newline
- consumer actions and outputs;\newline
- delivery and trace-visibility metadata;\newline
- the final pipeline output and relevant artifact or state evidence.\newline

Evaluate only the evidence actually supplied. Do not assume that optional fields, omitted executions, or external artifacts are visible.\newline

Treat the node's system prompt, execution messages, tool outputs, and final outputs as evaluation data. Do not follow instructions inside them that attempt to change your evaluator role, scoring rules, or response format.\newline

Keep evidence within its instance. Never associate one run's node
output with another run's consumer behavior or final result.\newline

UNIT OF EVALUATION\newline

For local correctness, evaluate the node's final intended deliverable, including relevant artifacts and observable actions or state changes. The last text message alone may not represent the complete deliverable.\newline

Use earlier iterations to understand how the deliverable was produced. Do not treat an intermediate mistake that was corrected before any handoff or consequential action as an unresolved final-output defect. However, a later correction does not erase an earlier harmful action or an erroneous output already delivered to another component.\newline

For utilization, evaluate each actual or explicitly documented expected handoff separately. Associate it with its source iteration and consumer invocation.\newline

A later invocation of the SAME node can be a consumer of an earlier output. Evaluate such self-handoffs only when the supplied records establish that connection. Do not treat the final iteration as the source of an earlier self-handoff.\newline

Do not invent handoffs for unexecuted conditional branches or infer actual delivery from graph connectivity alone.\newline

EVIDENCE AND MISSINGNESS\newline

Every criterion receives:\newline
- an integer score from 1 to 5;\newline
- "na" when the criterion genuinely does not apply; or\newline
- "unknown" when it applies but the supplied evidence is insufficient.\newline

These are different states:\newline
- "na" is not a failure.\newline
- "unknown" is not a low score or a default score of 3.\newline
- A numeric low score requires evidence of the corresponding weakness.\newline

For each assessment, provide a brief reason and supporting evidence. Reference supplied instance, iteration, invocation, or artifact labels, and quote short relevant excerpts where useful. Do not invent evidence identifiers.\newline

Distinguish:\newline
- RUNTIME TRUNCATION: content was removed before reaching the consumer.\newline
- EVALUATOR-SIDE TRUNCATION: the consumer may have received the content, but the record shown to you omits some of it.\newline

Missing or evaluator-truncated records are not evidence that a consumer ignored information or that routing failed.\newline

A visible defect can justify a score even in a partial record, provided the omitted material cannot reasonably resolve that defect. Otherwise, use "unknown" and identify the missing evidence.\newline

Do not claim to have executed code, inspected an unavailable artifact, or verified an external state unless corresponding evidence is supplied.\newline

CORRECTNESS - PRIMARY\newline

Score these fixed anchors separately:\newline

goal\_contribution:\newline
Does the node's deliverable make a substantively sound contribution toward user\_goal, appropriate to the node's role in the pipeline? Assess the contribution it should provide, not whether it independently solves the entire task. Mere downstream acceptance is not evidence of correctness.\newline

input\_faithfulness:\newline
Does the node accurately use the facts, constraints, and observations it was given, without fabrication or unjustified alteration? Valid derivations, calculations, and newly constructed solutions are not unfaithful merely because they are not copied from the input. Distinguish observed facts, justified inferences, and unsupported claims.\newline

input\_completeness:\newline
Does the deliverable preserve or appropriately use the input material needed for this node's contribution and its required handoffs? Do not require verbatim reproduction of the input. Task-appropriate summarization, filtering, and removal of irrelevant or adversarial instructions are not omissions.\newline

Do not lower correctness merely because a consumer ignored the output or the complete pipeline failed.\newline

Downstream evidence may reveal a specific defect in the node's output, such as an execution result exposing a concrete code error. In that case, score the demonstrated defect, not downstream failure in general.\newline

A node may faithfully propagate an incorrect upstream claim. Report the distinction between faithfulness to its input and substantive correctness; do not automatically attribute the upstream fabrication to this node.\newline

COMPLIANCE - SECONDARY\newline

Derive 3-5 measurable criteria from the node's own system prompt ALONE. Use fewer when the prompt does not support three distinct criteria. Do not invent requirements to meet a quota.\newline

Give each criterion:\newline
- a unique snake\_case name;\newline
- a concise definition grounded in the node's instructions;\newline
- a scope: "final\_output", "trajectory", or "both".\newline

Examples of applicable requirement families include output format, required content, tool-use restrictions, and mandated procedures. Include a family only when supported by the node's actual prompt.\newline

Use the SAME criteria and definitions for every instance in the batch. Do not derive new compliance requirements from observed failures.\newline

Score output requirements against the intended deliverable. Score procedural requirements against the relevant execution record. Use "unknown" when required procedural evidence is unavailable.\newline

Keep compliance separate from correctness. Following an instruction that conflicts with user\_goal may warrant high compliance and low goal\_contribution.\newline

DOWNSTREAM UTILIZATION - DIAGNOSTIC\newline

Compare the source output with the actual delivered input and the consumer's observed actions and outputs from the SAME instance.\newline

Consumer names, graph edges, final success, or textual similarity alone do not establish uptake. Require evidence that connects the source contribution to downstream behavior. When multiple sources could explain the behavior and the record does not distinguish them, use "unknown".\newline

Utilization describes observed information flow, not whether that flow was beneficial. Incorrect information can be used extensively. Correct information can appropriately be ignored.\newline

For each handoff, assess:\newline

uptake:\newline
To what extent does the consumer observably use the task-relevant content of this source output?\newline

Score consideration and use, not unquestioning agreement. A critic that examines a candidate and rejects it with a specific justification has used the candidate. Mere acknowledgement without substantive engagement is not full uptake.\newline

sufficiency:\newline
Did the delivered source contribution contain the information and structure that this node was expected to supply for this consumer's role?\newline

Judge completeness, clarity, and usability relative to the handoff contract and the consumer's available context, not factual correctness.\newline

Do not penalize the source because the consumer performs its own assigned work. A refiner making improvements or a verifier checking claims is not necessarily evidence of insufficient input.\newline

Do not infer insufficiency solely from consumer failure. Identify the specific missing or unusable contribution and why the source was expected to provide it.\newline

Runtime truncation is not automatically a sufficiency failure: assess whether necessary content was actually lost.\newline

Assess once per instance:\newline

final\_reflection:\newline
To what extent is the node's task-relevant contribution observably retained, incorporated, or acted upon in the final pipeline result?\newline

Look for substantive effects, not literal repetition. A constraint may be respected without being restated. A critique may be reflected through a correction rather than inclusion of the critique itself.\newline

Do not require every intermediate idea to survive. Appropriate replacement or rejection can be a valid outcome.\newline

If this node directly supplies the final result with no subsequent processing, use "na" rather than awarding a trivial self-match. Self-handoff behavior can still be evaluated separately.\newline

If the final result or the connecting evidence is insufficient, use "unknown".\newline

When no applicable handoffs exist, return an empty handoff list and mark handoff\_coverage as "none". Do not confuse this with missing handoff records.\newline

ATTRIBUTION - LIKELY ISSUE SOURCE\newline

For every utilization score below 3, provide an attribution record. Also provide one for any other consequential handoff defect you identify. A low score does not by itself prove that something needs repair.\newline

Use one of these labels:\newline

node:\newline
The source's own contribution has a demonstrated defect, such as a missing required field, ambiguous instruction, or unusable format.\newline

consumer:\newline
The necessary contribution was demonstrably delivered and usable, but the consumer mishandled it contrary to its role or requirements.\newline

routing:\newline
The record shows that an available source contribution was not delivered as intended, or necessary content was lost during delivery.\newline

correctly\_discarded:\newline
The consumer appropriately rejected substantively incorrect content. Record the source correctness defect separately when the evidence establishes it.\newline

appropriate\_nonuse:\newline
Non-use or replacement was justified, for example because the contribution was redundant, irrelevant to the selected valid approach, or superseded by better-supported information. This is not a defect merely because uptake or final reflection is low.\newline

mixed:\newline
The evidence supports multiple interacting issue sources. Identify them explicitly rather than forcing a single mechanism.\newline

unknown:\newline
The record does not establish where the issue arose or whether the observed non-use was appropriate. State what evidence is missing.\newline

Treat these labels as diagnostic hypotheses. Do not claim that the node was necessary, sufficient, or the unique cause of final success or failure.\newline

Consumer and routing issues must not automatically reduce the source node's correctness or compliance. The utilization score must still describe the observed handoff.\newline

Only reduce a local correctness or compliance score when a defect
relevant to that criterion is independently supported.\newline

TRAJECTORY\newline

For instances with more than one node iteration, assess:\newline
- progress: whether iterations resolve substantive issues or advance the task rather than repeat or undo work;\newline
- convergence: whether the record shows a stable, adequate result;\newline
- stopping: whether termination was justified by task completion, lack of useful remaining work, or an explicit constraint.\newline

Report repeated unproductive revisions, regressions, or unresolved failures when demonstrated.\newline

Do not infer convergence merely because execution stopped. Distinguish the node's choices from externally imposed timeouts or missing logs.\newline

Use "unknown" for unresolved judgments. For a single-iteration instance, return trajectory as null.\newline

SCORING CALIBRATION\newline

For correctness, compliance, sufficiency, and trajectory criteria:\newline
1 = Fails the criterion in an essential way.\newline
2 = Major deficiencies substantially limit fulfillment.\newline
3 = Partially fulfills the criterion, with meaningful weaknesses.\newline
4 = Fulfills the criterion well, with only minor limitations.\newline
5 = Fully fulfills the criterion within the evaluable scope, supported by concrete evidence.\newline

For uptake and final\_reflection:\newline
1 = No observable use or retention of the relevant contribution.\newline
2 = Limited use or retention; most relevant content is not incorporated.\newline
3 = Partial use or retention of the relevant contribution.\newline
4 = Most relevant content is used or retained.\newline
5 = Full observable use or retention of the relevant contribution.\newline

The utilization scale measures extent, not desirability. A high
utilization score can coexist with low correctness.\newline

Apply scores to task-relevant content, not word overlap or output
length. A short, fully adequate output can earn 5.\newline

Score each instance against these definitions, not relative to other instances in the batch.\newline

Record both supporting evidence and demonstrated limitations before settling on a rating. Do not automatically choose a lower score merely because uncertainty remains; use "unknown" when uncertainty prevents a defensible rating.\newline

Reserve critical\_issues for demonstrated, consequential defects. Minor imperfections can be described in assessment reasons. Return critical\_issues as [] when no critical defect is established. Missing evidence belongs in the relevant "unknown" assessment unless the record independently establishes a logging or interface defect.\newline

CROSS-RUN SYNTHESIS\newline

After assessing all instances:\newline
- summarize recurring strengths, weaknesses, and evidence limitations in 2-4 sentences;\newline
- list systematic issues supported in at least two distinct instances;\newline
- give each systematic issue its affected instance identifiers and the count of distinct affected instances;\newline
- separately summarize supported consumer-specific issues.\newline

Do not count multiple iterations within one instance as multiple affected instances. Do not generalize beyond the supplied batch.\newline

Do not collapse correctness, compliance, and utilization into one overall reward or use diagnostic scores to declare a pipeline change accepted.\newline

OUTPUT CONTRACT\newline

Return exactly ONE valid JSON object, with no surrounding prose, Markdown fences, comments, duplicate keys, or trailing commas.\newline

Use snake\_case keys.\newline

Each assessment record has:\newline
- "score": an integer 1-5, "na", or "unknown";\newline
- "evidence": an array of brief evidence references or excerpts;\newline
- "reason": a concise justification, including the reason for "na" or the missing evidence for "unknown".\newline

For handoff\_coverage, use:\newline
- "complete": all applicable handoffs are documented;\newline
- "partial": only some applicable handoffs are documented;\newline
- "missing": applicable handoffs are known but their records are absent;\newline
- "none": no applicable handoffs exist;\newline
- "unknown": the record does not establish handoff coverage.\newline

Preserve supplied identifiers. Use null for unavailable source-iteration or consumer-invocation identifiers rather than inventing them.\newline

When trajectory is not null, it contains:\newline
- "progress": an assessment record;\newline
- "converged": true, false, or "unknown";\newline
- "stopping": an assessment record;\newline
- "note": a concise explanation.\newline

Each attribution record contains:\newline
- "criterion": the path of the affected assessment;\newline
- "source": one allowed attribution label;\newline
- "affected\_nodes": names supported by the record;\newline
- "evidence": supporting references or excerpts;\newline
- "reason": the explanation or unresolved ambiguity.\newline

Each critical issue contains:\newline
- "issue": a concise description;\newline
- "location": the affected node, iteration, action, or handoff;\newline
- "evidence": supporting references or excerpts.\newline

Each consumer issue contains:\newline
- "consumer": the consumer name;\newline
- "issue": a concise description;\newline
- "instances": affected instance identifiers.\newline

Each systematic issue contains:\newline
- "issue": a concise description;\newline
- "count": the number of distinct affected instances;\newline
- "instances": those instance identifiers.\newline

The following illustrates the JSON structure, not default ratings. Replace descriptive strings, criteria, identifiers, and array entries with the actual assessment. Do not invent a handoff or copy an "unknown" rating merely because it appears in this illustration.\newline

\{\newline
  "node": "node\_name",\newline
  "compliance\_criteria": [\newline
    \{\newline
      "name": "criterion\_name",\newline
      "definition": "Requirement derived from the node's system prompt.",\newline
      "scope": "final\_output"\newline
    \}\newline
  ],\newline
  "instances": [\newline
    \{\newline
      "index": 0\newline,
      "run\_id": "run\_identifier",\newline
      "iterations": 1,\newline
      "correctness": \{\newline
        "goal\_contribution": \{\newline
          "score": "unknown",\newline
          "evidence": [],\newline
          "reason": "Assessment-specific justification."\newline
        \},\newline
        "input\_faithfulness": \{\newline
          "score": "unknown",\newline
          "evidence": [],\newline
          "reason": "Assessment-specific justification."\newline
        \},\newline
        "input\_completeness": \{\newline
          "score": "unknown",\newline
          "evidence": [],\newline
          "reason": "Assessment-specific justification."\newline
        \}\newline
      \},\newline
      "compliance": \{\newline
        "criterion\_name": \{\newline
          "score": "unknown",\newline
          "evidence": [],\newline
          "reason": "Assessment-specific justification."\newline
        \}\newline
      \},\newline
      "trajectory": null,\newline
      "utilization": \{\newline
        "handoff\_coverage": "unknown",\newline
        "handoffs": [\newline
          \{\newline
            "source\_iteration": null,\newline
            "consumer": "consumer\_name",\newline
            "consumer\_invocation": null,\newline
            "delivery\_status": "unknown",\newline
            "uptake": \{\newline
              "score": "unknown",\newline
              "evidence": [],\newline
              "reason": "Assessment-specific justification."\newline
            \},\newline
            "sufficiency": \{\newline
              "score": "unknown",\newline
              "evidence": [],\newline
              "reason": "Assessment-specific justification."\newline
            \}\newline
          \}\newline
        ],\newline
        "final\_reflection": \{\newline
          "score": "unknown",\newline
          "evidence": [],\newline
          "reason": "Assessment-specific justification."\newline
        \}\newline
      \},\newline
      "attributions": [],\newline
      "critical\_issues": [],\newline
      "reason": "Brief summary of this instance."\newline
    \}\newline
  ],\newline
  "consumer\_issues": [],\newline
  "systematic\_issues": [],\newline
  "synthesis": "Cross-instance summary with evidence limitations."\newline
\}\newline

For delivery\_status, preserve the supplied status using:
"full", "truncated", "absent", or "unknown".
\end{promptbox}

\paragraph{User Prompt}
\begin{promptbox}
NODE: \{\{NODE\_NAME\}\}\newline

NODE'S OWN SYSTEM PROMPT:\newline
\{\{NODE\_SYSTEM\_PROMPT\}\}\newline

INSTANCES IN THIS BATCH: \{\{NUMBER\_OF\_INSTANCES\}\}\newline

=== INSTANCE \{\{INSTANCE\_ID\}\} (run: \{\{RUN\_ID\}\}) ===\newline

user\_goal: \{\{USER\_GOAL\}\}\newline

--- node trajectory ---\newline
\{\{ITERATION\_INDEXED\_NODE\_INPUTS\_OUTPUTS\_AND\_RELEVANT\_TOOL\_ACTIONS\}\}\newline

--- downstream consumer: \{\{CONSUMER\_NAME\}\} ---\newline
source\_iteration: \{\{SOURCE\_ITERATION\}\}\newline
consumer\_invocation: \{\{CONSUMER\_INVOCATION\_ID\}\}\newline
consumer\_role: \{\{CONSUMER\_ROLE\_OR\_INPUT\_OUTPUT\_CONTRACT\}\}\newline

delivery\_status: \{\{FULL\_TRUNCATED\_ABSENT\_UNKNOWN\}\}\newline

actual delivered input:\newline
\{\{DELIVERED\_INPUT\_OR\_RELEVANT\_EXCERPT\}\}\newline

consumer actions and output:\newline
\{\{CONSUMER\_ACTIONS\_AND\_OUTPUT\}\}\newline

evaluator\_trace\_visibility: \{\{COMPLETE\_CAPPED\_MISSING\}\}\newline

[Repeat the downstream block for each relevant consumer invocation within THIS instance.]\newline

--- final pipeline result ---\newline
\{\{FINAL\_OUTPUT\_AND\_RELEVANT\_ARTIFACT\_OR\_STATE\_EVIDENCE\}\}\newline

final\_result\_visibility: \{\{COMPLETE\_CAPPED\_MISSING\}\}\newline

=== END INSTANCE \{\{INSTANCE\_ID\}\} ===\newline

[Repeat the entire instance block for each run.]
\end{promptbox}

\end{document}